\documentclass[namedreferences]{SolarPhysics}
\usepackage[optionalrh]{spr-sola-addons} 
\usepackage{graphicx}
\usepackage{url}             

\begin{document}
\begin{article}
\begin{opening}

\title{Absorption Phenomena and a Probable Blast Wave in the 13 July 2004 Eruptive Event}

\author{V.V.~\surname{Grechnev}$^{1}$\sep
        A.M.~\surname{Uralov}$^{1}$\sep
        V.A.~\surname{Slemzin}$^{2}$\sep
        I.M.~\surname{Chertok}$^{3}$\sep
        I.V.~\surname{Kuzmenko}$^{4}$\sep
        K.~\surname{Shibasaki}$^{5}$}

\runningauthor{Grechnev et al.} \runningtitle{Absorption and Blast
Wave}

\institute{$^{1}$ Institute of Solar-Terrestrial Physics SB RAS,
Lermontov St.\ 126, Irkutsk 664033, Russia
                  email: \url{grechnev@iszf.irk.ru} email: \url{uralov@iszf.irk.ru}\\
           $^{2}$ P.N. Lebedev Physical Institute, Lenisky Pr., 53, Moscow, 119991,
                  Russia
                  email: \url{slem@lebedev.ru}\\
           $^{3}$ Pushkov Institute of Terrestrial Magnetism,
                  Ionosphere and Radio Wave Propagation (IZMIRAN), Troitsk, Moscow
                  Region, 142190 Russia
                  email: \url{ichertok@izmiran.ru}\\
           $^{4}$ Ussuriysk Astrophysical Observatory, Solnechnaya St. 21, Primorsky
                  Krai, Gornotaezhnoe 692533, Russia
                  email: \url{kuzmenko_irina@mail.ru}\\
           $^{5}$ Nobeyama Radio Observatory, Minamimaki, Minamisaku,
                  Nagano 384-1305, Japan
                  email: \url{shibasaki@nro.nao.ac.jp}}

\date{Received ; accepted }

\begin{abstract}
We present a case study of the 13 July 2004 solar event, in which
disturbances caused by eruption of a filament from an active
region embraced a quarter of the visible solar surface. Remarkable
are absorption phenomena observed in the SOHO/EIT 304~\AA\
channel; they were also visible in the EIT 195~\AA\ channel, in
the H$\alpha$ line, and even in total radio flux records. Coronal
and Moreton waves were also observed. Multi-spectral data allowed
reconstructing an overall picture of the event. An explosive
filament eruption and related impulsive flare produced a CME and
blast shock, both of which decelerated and propagated
independently. Coronal and Moreton waves were kinematically close
and both decelerated in accordance with an expected motion of the
coronal blast shock. The CME did not resemble a classical
three-component structure, probably, because some part of the
ejected mass fell back onto the Sun. Quantitative evaluations from
different observations provide close estimates of the falling
mass, $\sim 3 \cdot 10^{15}\,$g, which is close to the estimated
mass of the CME. The falling material was responsible for the
observed large-scale absorption phenomena, in particular, shallow
widespread moving dimmings observed at 195~\AA. By contrast, deep
quasi-stationary dimmings observed in this band near the eruption
center were due to plasma density decrease in coronal structures.
\end{abstract}

\end{opening}

\section{Introduction}

Some major flares and coronal mass ejections (CMEs) are
accompanied by large-scale disturbances moving over distances
comparable with the solar radius like Moreton waves, weakly bright
coronal transients (Thompson \textit{et al.}, 1998) usually termed
`EIT waves', and temporary depressions of soft X-ray (SXR) and
extreme-ultraviolet (EUV) emissions (termed dimmings) appearing
near eruption centers and far from them. They have been observed
with the Extreme-ultraviolet Imaging Telescope (EIT;
Delaboudini\`ere \textit{et al.}, 1995) on SOHO, the Transition
Region and Coronal Explorer (TRACE; Handy \textit{et al.}, 1999),
and other telescopes. Nevertheless, nature of all these phenomena
is still controversial.

Moreton waves are sometimes observed in the H$\alpha$ line
(Moreton, 1960) and, more recently, in the He{\sc i} 10830~\AA\
line (\textit{e.g.}, Vr{\v s}nak \textit{et al.}, 2002; Gilbert
\textit{et al.}, 2004a). Uchida (1968) proposed them to be due to
coronal disturbances. Balasubramaniam, Pevtsov, and Neidig (2007)
also argued their coronal nature. Moreton waves run faster
(400--2000~km\thinspace s$^{-1}$) than `EIT waves' ($<
800$~km\thinspace s$^{-1}$, typically $\simeq\,$250~km\thinspace
s$^{-1}$). Warmuth \textit{et al.} (2001, 2004a, 2004b, 2005)
found their kinematical closeness and proposed that both
chromospheric and coronal signatures of the waves could be created
by a single decelerating disturbance, \textit{e.g.}, a blast shock
(see also Hudson and Warmuth, 2004). Gilbert \textit{et al.}
(2004a, 2004b) found `EIT waves' to be co-spatial with their
counterparts in He{\sc i}.

Some observations favor the wave nature of `EIT waves' (Ballai,
Erd{\' e}lyi, and Pint{\' e}r, 2005). A few moving bright features
were interpreted as CME frontal magnetic structures (Dere
\textit{et al.}, 1997; Uralov, Grechnev, and Hudson, 2005).
Several authors propose explanations for `EIT waves' within
scenarios suggested for dimmings. Delann{\'e}e and Aulanier (1999)
proposed `EIT waves' to be running boundaries of successively
opening magnetic structures ahead of expanding dimming areas, from
which plasmas evacuate. Attrill \textit{et al.} (2007) proposed to
explain `EIT waves' along with expanding shallow dimmings in their
scenario based on interchange reconnection. However, `EIT waves'
are not commonly associated with dimmings (\textit{e.g.}, Chertok
and Grechnev, 2005b; Grechnev \textit{et al.}, 2005). This means
that their nature is not always the same.

Trying to reconcile contradictory views, Zhukov and Auch{\` e}re
(2004) proposed that `EIT waves' represent both wave-like and
eruptive components. Simulations of Chen \textit{et al.} (2002,
2005) confirmed a possibility of two types of coronal wave
phenomena during an eruption, one according to the scenario of
Delann{\'e}e and Aulanier (1999), and a fast coronal counterpart
of the Moreton wave ahead.

Quasi-stationary dimmings develop from $\sim$10 minutes to several
hours, reach $-80\%$ (drop up to $20\%$ of the pre-event
brightness; Chertok and Grechnev, 2003a), and live up to two days
(Hudson and Webb, 1997). Having once appeared in some region, they
persist there (Chertok and Grechnev, 2005b; Grechnev et al.,
2005). Such dimmings are related with active regions and
large-scale magnetic fields (Chertok and Grechnev, 2005b; Slemzin,
Kuzin, and Bogachev, 2005; Slemzin, Grechnev, and Kuzin, 2006;
Zhang \textit{et al.}, 2007). Their main interpretation is plasma
density depletion due to its outflow from previously closed
structures (see also Harra and Sterling, 2001; Harra \textit{et
al.}, 2007).

Some moving shallow darkenings seem to accompany `EIT waves' in
their expansion (see, \textit{e.g.}, Zhukov and Auch{\` e}re,
2004). They can have the same wave-like nature as such `EIT
waves'. Another kind of short-lived, moving darkenings could be
due to absorption of the background emission by material of
ejecta. Such phenomena have been observed in images produced with
EIT in the 195~\AA\ (Fe\thinspace {\sc xii}) band and TRACE in the
195 and 173~\AA\ (Fe\thinspace {\sc ix}) bands (the 173 channel is
usually called 171 to be consistent with the terminology of EIT;
we use 173 to distinguish them). In particular, darkenings
sometimes observed with EIT in the 304~\AA\ band (mainly
He\thinspace {\sc ii}, 20\,000--80\,000~K) without counterparts in
coronal bands (Chertok and Grechnev, 2003b) are expected to be
certainly due to absorption in the ejected filament material
(Delaboudini\`ere, 2005). Such a phenomenon was recorded with the
CORONAS-F/SPIR\-IT EUV telescope (Zhitnik \textit{et al.}, 2002)
in a major event of 18 November 2003 (Slemzin \textit{et al.},
2004; Grechnev \textit{et al.}, 2005). A dark feature as large as
the solar radius was observed at 304~\AA\ only to move during
1.5~hr across the Sun with the plane-of-sky speed of $\simeq
200\,$km\thinspace s$^{-1}$. Thus, the increasing observational
material suggests that different kinds of `EIT waves' and dimmings
probably exist (as also Warmuth \textit{et al.} (2001) proposed).

`Negative bursts' (transient decreases of the total radio flux
below the relatively quiet level of the solar emission) are
believed to be mainly also due to absorption (\textit{e.g.},
Covington and Dodson, 1953; Maksimov and Nefedyev, 1991), although
other reasons are possible (\textit{e.g.}, Sawyer, 1977). Since
opacity of an absorber depends on frequency and plasma parameters,
multi-frequency observations of such bursts promise quantitative
diagnostics of ejecta. It seems to be useful to compare
observations of an event, in which absorption occurs in both EUV
and radio ranges.

This paper addresses the 13 July 2004 event, in which a
multi-component eruption, CME, Moreton and coronal waves,
large-scale EUV absorption phenomena, long-lived dimmings, and a
`negative radio burst' were observed simultaneously. The event
included an impulsive M6.7 flare (00:09--00:23, \textit{all times
hereafter are UT}) in active region 10646 (N13\thinspace W46) and
a CME with a central position angle of $294^{\circ}$ observed
after 00:54 with the Large-Angle Spectrometric Coronagraph (LASCO;
Brueckner \textit{et al.}, 1995) on SOHO. EIT observed large-scale
disturbances in the 195~\AA\ band, in particular, a faint oval
front at 00:24. A type II burst at 00:16--00:43 was reported by a
few observatories. At 01:19, a large-scale darkening was observed
with EIT at 304~\AA. During this event, a strong impulsive radio
burst was recorded at 1--80~GHz followed by a decrease of the
radio emission. The `negative burst' was revealed in total flux
records of the Ussuriysk Astrophysical Observatory (UAFO, 2.8
GHz), Nobeyama Radio Polarimeters (NoRP; Nakajima \textit{et al.},
1985; Torii \textit{et al.}, 1979), and the Learmonth Observatory.
The event was also recorded in the H$\alpha$ central line in the
Big Bear (BBSO) and Mauna Loa (MLSO) Solar Observatories, in the
He{\sc i} line in MLSO, and in the TRACE 173~\AA\ channel. The
Nobeyama Radioheliograph (NoRH; Nakajima \textit{et al.}, 1994)
observed some fragments of the ejecta at 17~GHz.

Using multi-spectral data, we study this event and reconstruct a
picture of the eruption. The ejecta disintegrated. One part fell
back onto the Sun, while another part escaped. Dispersed material
was responsible for the absorption phenomena observed. Estimations
of masses of both these parts, performed using different methods,
quantitatively confirm this picture.

The event started with an explosive filament eruption and
impulsive flare that likely produced a blast wave. Kinematics of
the wave and ejected fragments are compared with analytical
expressions, which match observations. Other aspects of the event
concerning radio bursts, pulsations, and their association with
the flare, ejecta, and shocks are addressed by Pohjolainen, Hori,
and Sakurai (2008).

Section~\ref{Observations} presents observational data. All
measurements are related to the plane of sky (POS), if otherwise
not specified. In Section~\ref{Absorption}, we address absorption
in different emissions and estimate the mass of the ejecta. In
Section~\ref{Kinematics}, we discuss kinematical characteristics
of the observed phenomena and compare them with theoretical
expectations. In Section~\ref{Discussion}, we discuss results of
our analysis, the overall picture of the event, and nature of `EIT
waves' and dimmings in this event. Section~\ref{Summary}
summarizes our findings and their implications.

\section{Observations}
\label{Observations}

\subsection{Filament Eruption, Moreton Wave, and Surge in H$\alpha$ Images}
\label{H_alpha}

\begin{figure}    
   \centerline{\includegraphics[width=\textwidth,clip=]{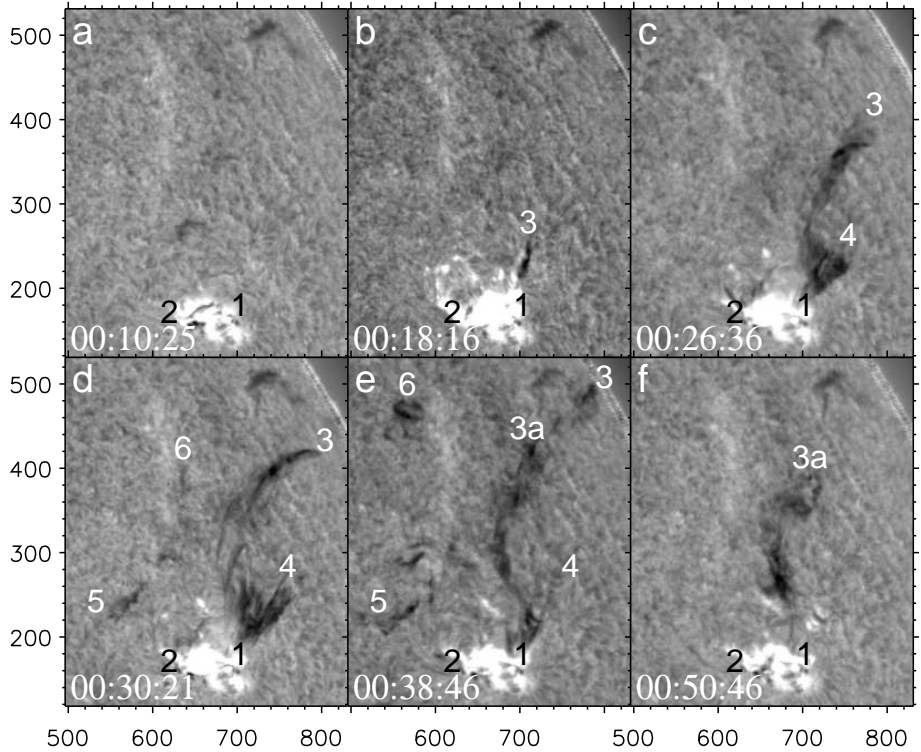}
              }
              \caption{
H$\alpha$ images produced in the Big Bear Solar Observatory. Limb
darkening removed. Labels denote features discussed in the text.
Axes hereafter show arc seconds from the solar disk center.
              }
 \label{Big_Bear_H-alpha}
   \end{figure}

\begin{figure}    
   \centerline{\includegraphics[width=\textwidth,clip=]{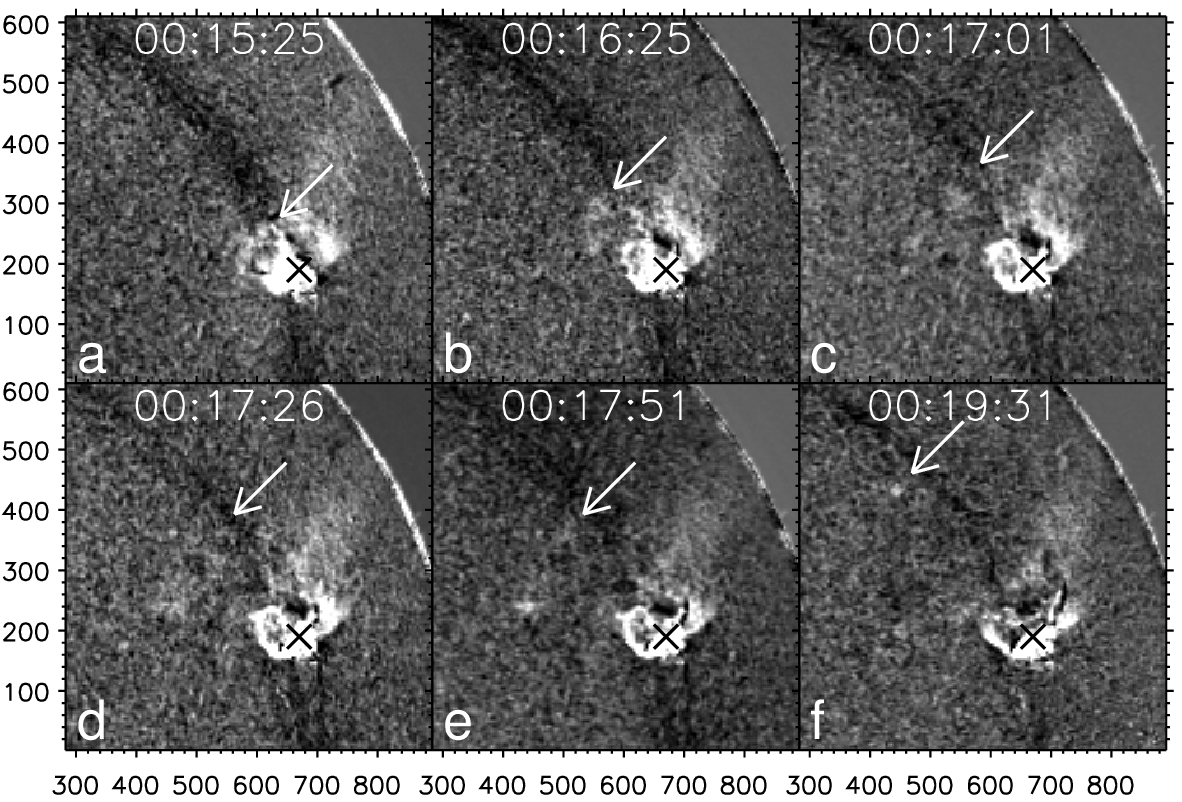}
              }
              \caption{
Moreton wave (the front indicated by the arrows) in BBSO H$\alpha$
images with subtracted average image. The dark wide ring is an
artifact (a `trace' of the active region that appeared in a
toroidal Fourier transform in the course of background
subtraction). All images are displayed nonlinearly to emphasize
the wave. The cross marks the eruption center.
              }
 \label{Moreton_wave}
   \end{figure}

The 13 July 2004 event was observed in the H$\alpha$ line center
in BBSO and MLSO. No H$\alpha$ flare was reported for this event,
but images from both observatories show it and related eruptive
activities. Figure~\ref{Big_Bear_H-alpha} presents H$\alpha$
images obtained in BBSO. The electronic version of our paper
contains a movie 2004-07-13\_BBSO\_Ha.mpg, whose left panel shows
H$\alpha$ images with a limb darkening removed, and the right
panel shows the same images after additional processing to reveal
features of interest. Before the eruption, the active region
contained a system of rather small filaments. Filament located
between positions 1 and 2 starts to rise at 00:02:30 with the top
part accelerating at $\simeq 16\,$m\thinspace s$^{-2}$, which
reaches 7~Mm at 00:13:25 with a speed of $\gsim 22\,$km\thinspace
s$^{-1}$ ($2^\mathrm{nd}$ order fit). It erupts between 00:13:25
and 00:15:25. A Moreton wave runs from the eruption site
north/northeast (see Figure~\ref{Moreton_wave}). At about its
onset, surge 3 starts to grow from the western end 1 of the
filament with the average speed of $\simeq 270\,$km\thinspace
s$^{-1}$ (00:18--00:29). At 00:17\,--\,00:18, a surge from its end
2 appears and takes a V-like shape by 00:26 (frame \textit{c}).

The western surge 1\,--\,3 consisting of a multitude of thin,
likely twisted, threads intensively develops. After 00:22, it
starts to bifurcate into the northwest (1\,--\,4) and north
(1\,--\,3) branches. Branch 1\,--\,4 rises up to its maximum POS
height of $\simeq\,$90 Mm by 00:30 and then mainly disappears.
Branch 1\,--\,3 rises and slightly displaces east. New threads
appear. Along with fragments emanating from the eastern vicinity 2
of the main filament and a new remote feature 5, they form to
00:30 a large-scale filamentary structure. Its components occupy a
significant area 1--4--3--6--5--2 in frame \textit{d}. Then this
area continues to expand north and east, especially, due to the
northern part of branch 3 and fragment 6 (frame \textit{e}).
Feature 6 moves toward the North Pole with a speed $\simeq
155$~km\thinspace s$^{-1}$ (00:31--00:53). As MLSO He{\sc
i}~10830~\AA\ images show, the disturbance probably associated
with this moving feature reaches vicinities of the polar coronal
hole at $\simeq\,$01:20. At about 00:40, portion 3a of surge 1--3
overturns at a POS height of 190~Mm and starts to fall, while its
farthest part 3 keeps on moving to the limb. After 00:50, the
surge partially falls back and becomes fainter. This is due to
either spreading out its components that decreases its optical
depth, or their increasing line-of-sight velocities that results
in the Doppler brightening. Throughout development of the surge,
many threads and fragments keep connection with the position of
the initial filament(s) 1--2, and their bases rest on its (their)
vicinities.

From H$\alpha$ data we conclude that a filament eruption resulted
in formation of some large-scale `dome' of filament's threads and
fragments to cover in several tens of minutes a significant
northwest part of the Sun.

\subsection{High-Resolution TRACE 173~\AA\ Images of the Eruption}

\begin{figure}    
   \centerline{\includegraphics[width=\textwidth,clip=]{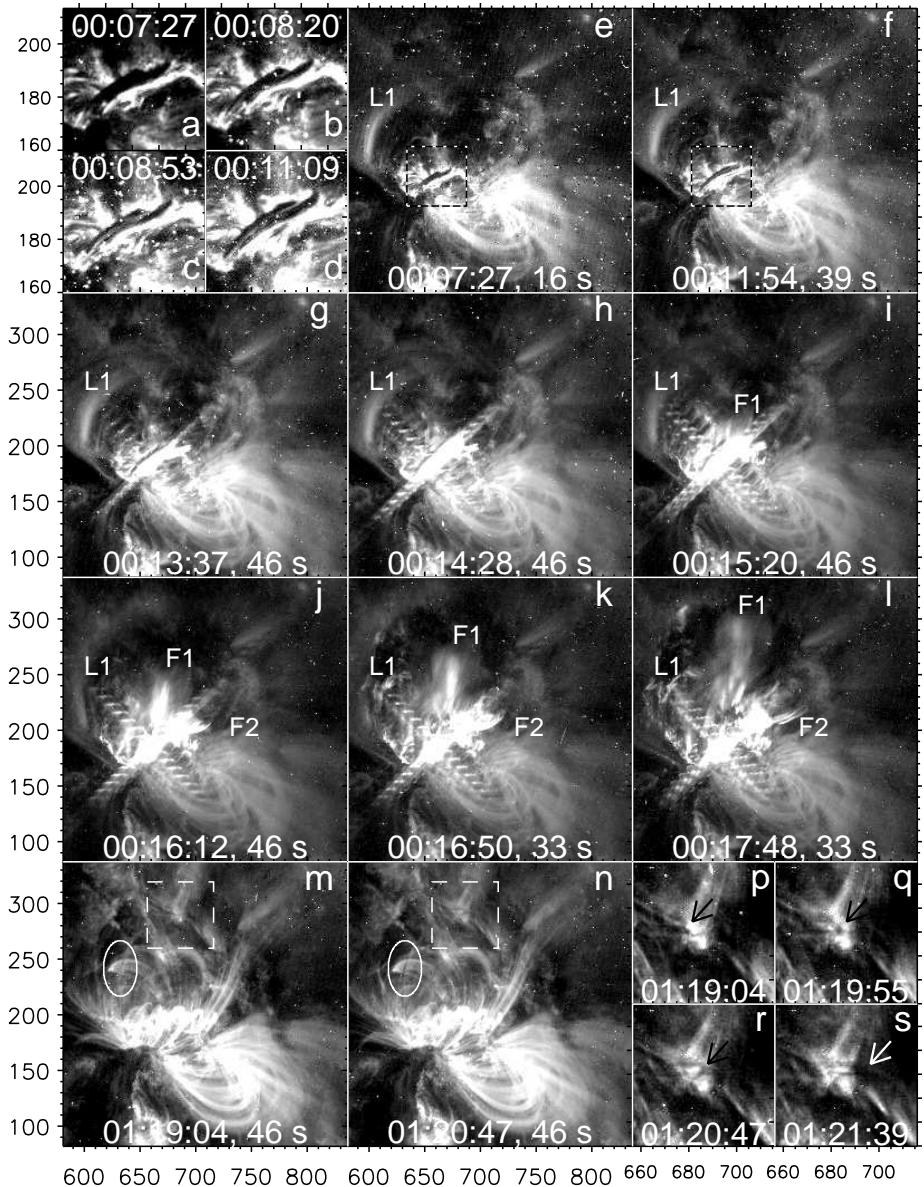}
              }
              \caption{
TRACE 173~\AA\ images of the eruption: \textit{a}--\textit{l}, the
early stage; \textit{m}--\textit{s}, the late stage. Insets
\textit{a}--\textit{d} show enlarged images of filaments outlined
in panels \textit{e} and \textit{f} by dotted frames. Labels
\textit{L1} in panels \textit{e}--\textit{l} mark an initial
position of an erupting loop. Labels \textit{F1} and \textit{F2}
mark bright jet-like moving features (position of label
\textit{F2} is fixed). Insets \textit{p}--\textit{s} show enlarged
areas outlined in panels \textit{m} and \textit{n} by dashed
frames, in which a flow of absorbing material is indicated by the
arrows. Another flowing absorbing feature is outlined by the
ellipse in panels \textit{m} and \textit{n}. Observation times and
exposure durations are specified in frames \textit{e}--\textit{n}.
All images are nonlinearly displayed. }
 \label{TRACE}
   \end{figure}

TRACE observed the event starting from 00:07:27 in the 173~\AA\
channel mainly with intervals from 20 to 70~s, except for the gap
of 00:21--00:48. Figure~\ref{TRACE} presents several images of the
eruption region, which TRACE permits to discern like a magnifier
(see also movie 2004-07-13\_TRACE\_173.mpg). Pre-eruption images
show a dark filament(s) and a bright rope along it as well as the
northern and southern loop systems surrounding the region.

Four enlarged images (this area is outlined in panels \textit{e}
and \textit{f} by dotted frames) in panels \textit{a}--\textit{d}
show two filaments located side-by-side to `merge' and then to
expand. The expansion of the combined filament is visible in
frames \textit{e}--\textit{l}. The rising filament gets
semi-transparent, and its parts become detectable
(\textit{j},\textit{k}). Loop \textit{L1} expands in frames
\textit{i}--\textit{k} and disappears by 00:18 (\textit{l}). A
flare starts beneath the filaments by 00:13:37. In frame
\textit{i}, bright, apparently diffuse material appears above the
whole combined filament as its upper envelope, and later, in
frames \textit{j}--\textit{l}, its brightest part appears as a
jet-like feature \textit{F1} moving north (up). Another, the
slower bright jet-like feature \textit{F2} moves west-northwest in
frames \textit{j}--\textit{l}. Crossly dark/bright patterns are
due to interference on the CCD matrix.

The POS speed of feature \textit{F1}, $v = 400-500$~km\thinspace
s${^{-1}}$, is acquired in $\leq 2$~min (acceleration $\geq
8$~km\thinspace s${^{-2}}$). With this speed and exposure times
$t_\mathrm{exp}$ specified in frames, apparent extents of this
feature are $v
t_\mathrm{exp}R_\odot^{\mathrm{arcsec}}/R_\odot^{\mathrm{km}}
\simeq 18^{\prime \prime}-31^{\prime \prime}$. Its jet-like
appearance is caused by blurring due to its displacement during
the exposure. The same reason explains the semi-transparent
appearance of the combined filament. It remains dark in all
frames, where it is visible (by 00:17:48). The speed of its top is
$\simeq (0.5-0.7) v$.

Hence, the eruption occurs between 00:14:28 and 00:15:20. The
eruptive filament remains cool. It is surrounded by a hotter
envelope with probable temperatures of 0.3\,--\,1.6~MK (the
temperature sensitivity range of the 173~\AA\ channel).

Expansion of the filament and adjacent structures starts before a
motion of loop \textit{L1}. Its displacement visible in frames
\textit{i,j}, when the ejecta was far from the loop, indicates
passage of a wave. Thus, loop \textit{L1} is pushed by the wave
rather than vice versa. The first manifestation of the wave in
frame \textit{i} is somewhat ahead of the Moreton wave
(Figure~\ref{Moreton_wave}a). The speed of the loop is $\simeq
350$~km\thinspace s$^{-1}$.

Images obtained after the gap in observations show a post-eruptive
arcade. Interesting are dark moving flows occulting brighter
structures behind them. Examples are outlined by dashed frames in
panels \textit{m,n} (insets \textit{p}--\textit{s} show enlarged
images, where a flow is indicated by the arrows) and by ovals.
These flows represent moving absorbing (probably cool) material
directed likely along magnetic field lines. Note also shrinkage of
loops that we do not discuss here.

The eruption destroyed the northern loop system. Loop \textit{L1}
erupted, but not reappeared afterwards. The southern loop system
remained almost unchanged. The eruption looks like a `directional
explosion' pointed north.

\subsection{Coronal Disturbances in SOHO/EIT 195~\AA\ Images}
\label{Disturbances}

\begin{figure}    
   \centerline{\includegraphics[height=17.cm,clip=]{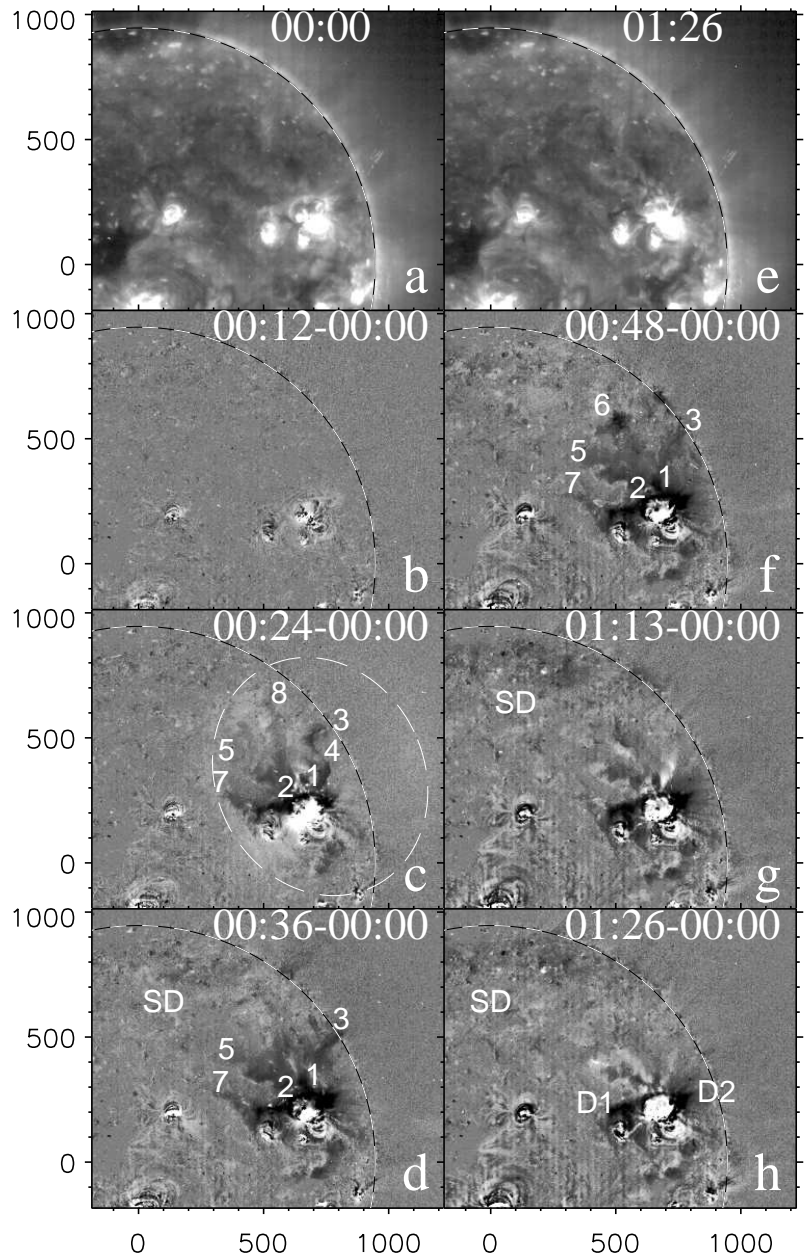}
              }
              \caption{
SOHO/EIT 195~\AA\ images of the event. (a)~Pre-event image,
(e)~later-stage image. Fixed-base (00:40~UT) differences (b--d and
f--h) show large-scale coronal disturbances. Nonlinear display is
applied to emphasize them. Dashed black-white circles denotes the
solar limb. Dashed white oval traces the aureole discussed in the
text. Labels 1--8 denote darkenings; D1 and D2, long-lived
dimmings.              }
 \label{EIT195_fixed_base_dif}
   \end{figure}

\begin{figure}    
   \centerline{\includegraphics[width=\textwidth,clip=]{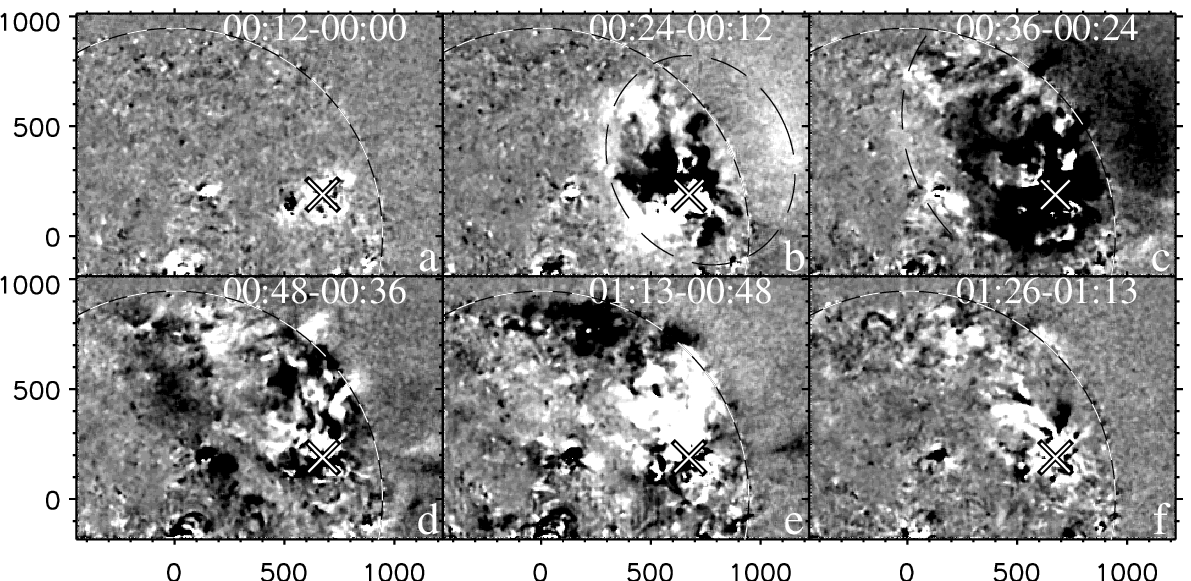}
              }
              \caption{
Large-scale disturbances in the NW quadrant visible in
running-difference SOHO/EIT 195~\AA\ images (nonlinearly
displayed). The cross marks the eruption center.              }
 \label{running_dif}
   \end{figure}

EIT produced 195~\AA\ images every 12 min (CME Watch program).
Figure~\ref{EIT195_fixed_base_dif} shows a pre-event (\textit{a})
image and a late-stage (\textit{e}) one along with six fixed-base
differences \textit{b\,--\,d} and \textit{f\,--\,h} with the solar
rotation preliminarily compensated (see Chertok and Grechnev,
2005b). Figure~\ref{running_dif} provides additional information
about large-scale faint disturbances, which are easier to see in
running differences.

The eruption causes significant disturbances in the corona during
00:24--01:26, both brightening and darkening, that cover a huge
area exceeding the NW quadrant of the solar disk. A nearly perfect
oval faint brightening is seen at 00:24
(Figures~\ref{EIT195_fixed_base_dif}c and \ref{running_dif}b). It
is not uniform in brightness, mainly 5--15 counts/pixel, while the
brightness in quiet Sun's regions is about 40--50 and 5--21 in
coronal holes. Three wide brightening sectors directed
east-southeast, northeast, and west-southwest spread toward the
oval from inside. The oval expands; its next position on the solar
disk is denoted by a dashed arc in Figure~\ref{running_dif}c
(00:36).

Darkening areas are pronounced in
Figure~\ref{EIT195_fixed_base_dif} (c,d, f--h). Some of them,
mostly short-lived once, coincide with dark features like surges
visible in H$\alpha$ images (\textit{1--6}, the same numbering).
Some others are different, \textit{e.g.}, \textit{7} and
\textit{8} and, especially, deep, long-lived dimmings. They show
up as two regions \textit{D1} and \textit{D2} near the
post-eruptive arcade, a dark arch between their tops, and a V-like
dark feature on top of the arch. Comparison of panels \textit{h},
\textit{a}, and \textit{e} shows that these deep, long-lived
dimmings are due to significant darkening or disappearance of
rather compact coronal structures, which were previously bright.
Dimmings of shorter lifetimes with counterparts in H$\alpha$
images are probably due to absorption of the 195~\AA\ emission in
the ejected material. A shallow depression (SD) moving toward the
North Pole is faintly visible in fixed-base difference images
(Figure~\ref{EIT195_fixed_base_dif}d,g,h).

\begin{figure}    
   \centerline{\includegraphics[width=\textwidth,clip=]{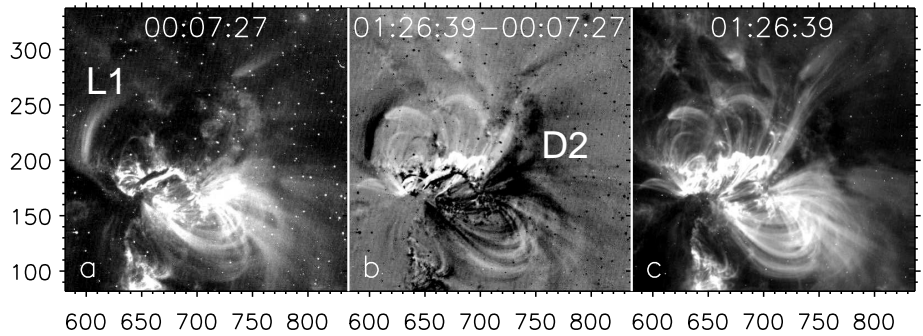}
              }
              \caption{
TRACE 173 \AA\ images before (a), after (c) the eruption and their
difference (b).              }
 \label{TRACE_dimming}
   \end{figure}

The field of view of TRACE is insufficient to cover dimmings
\textit{D1} and \textit{D2}. TRACE 173~\AA\ images in
Figure~\ref{TRACE_dimming} show that the arch-like dimming is due
to eruption of loop \textit{L1}. Dimming \textit{D2} is due to
changes of the loop system near the western leg of the filament.
The TRACE 173 \AA\ and EIT 195 \AA\ bands are different that
explains some dissimilarity of their images.

\subsection{Large-Scale Darkening in the SOHO/EIT 304~\AA\ Channel}

\begin{figure}    
   \centerline{\includegraphics[width=\textwidth,clip=]{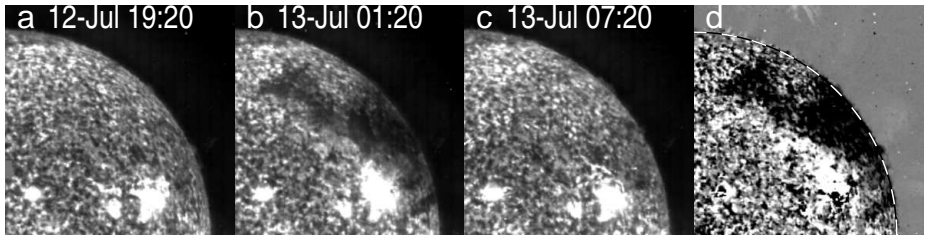}
              }
              \caption{
The northwest quadrant of EIT 304~\AA\ images produced before
(\textit{a}), during (\textit{b}), and after the event
(\textit{c}); (\textit{d}) difference image, \textit{frame (b)
minus frame (a)}. Solar rotation compensated for all images to
00:40.               }
 \label{EIT_304}
   \end{figure}

SOHO/EIT observed a large-scale darkening at 304~\AA\ in a single
frame at 01:20. Figure~\ref{EIT_304}a--c shows three sequential
full-disk images produced in the 304~\AA\ channel. They are
separated by 6 hours that rules out a possibility to study
development and motion of the dark feature. It is certainly absent
in the first (\textit{a}) and last (\textit{c}) frames, being well
pronounced in frame \textit{b}. Figure~\ref{EIT_304}d shows a
difference of frames \textit{b} and \textit{a}. Both positive and
negative pixel values are restricted. The area bounded by the
$-25\%$ level is 6.7\% of the solar disk, and the deepest
depression reaches $-60\%$. The configuration of the large-scale
darkening at 304~\AA\ is completely different from those observed
in the 195~\AA\ band.

\subsection{Ejecta Observed with NoRH at 17 GHz}

\begin{figure}    
   \centerline{\includegraphics[width=\textwidth,clip=]{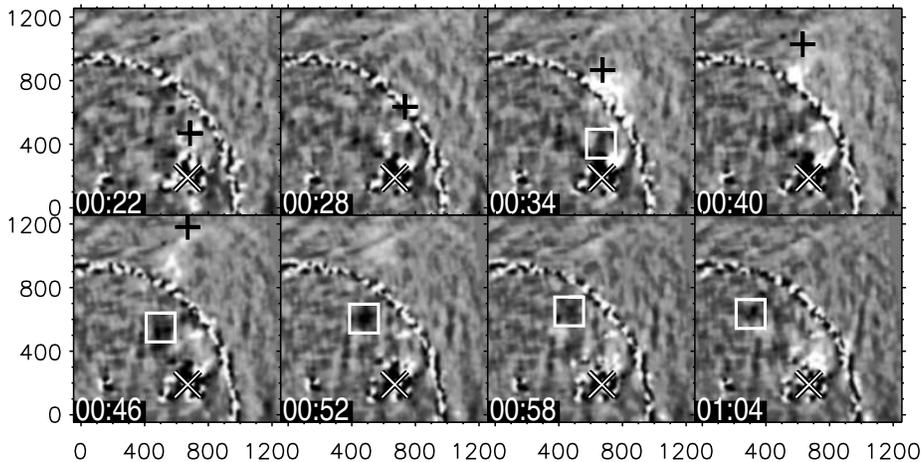}
              }
              \caption{
NoRH 17 GHz difference images (nonlinearly displayed). Black
crosses mark the measured leading edge of the bright feature.
White squares mark the measured center of the dark feature.
Slanted cross marks the eruption center.               }
 \label{NoRH_images}
   \end{figure}

The flare produced a strong microwave burst at 17~GHz, with its
main phase lasting until 00:19 and a maximum brightness
temperature up to $T_\mathrm{B} = 1.2 \cdot 10^8\,$K. To reveal
faint moving features, we produced enhanced-sensitivity NoRH
images at 17 GHz with an integration time of 1~min.
Figure~\ref{NoRH_images} shows some of them. Moving features are
still poorly visible, especially by cessation of the flare. Using
extra averaging over three frames and subtraction of an image
averaged during 01:41--01:50, it is possible to detect two moving
features in 17~GHz images.

One of them is dark and moves toward the North Pole (white squares
mark its center). Its lowest brightness temperature reaches
6\,000--8\,000~K against the quiet Sun's background (10\,000~K at
17~GHz) pointing at its large optical thickness and, consequently,
the kinetic temperature of $\lsim 6000\,$K. Another feature is
bright and moves up in the plane of sky. Black crosses mark its
roughly measured leading edge. It becomes pronounced after leaving
the solar disk. This is probably due to its overlap with the dark
feature, when both of them are not far from the eruption region.
Thus, this feature is probably optically thin. Its brightness
temperatures are 400--5\,000~K; hence, its kinetic temperature is
probably $\gg\,$5\,000~K. Its estimated velocity is $\sim
500$~km\thinspace s$^{-1}$, close to the speed of the bright
feature \textit{F1} visible in TRACE 173~\AA\ images. Their
trajectories are also close (see Figure~\ref{Scheme}). The bright
NoRH feature resembles in shape the foremost bright envelope
observed by TRACE at 00:15:20 (see Figure~\ref{TRACE}i). Although
the NoRH feature is not detectable during TRACE observations
because of bright flare emission, these facts demonstrate the NoRH
and TRACE bright features to be identical (some dissimilarities
are due to their different temperature responses).

\subsection{CME in SOHO/LASCO/C2 Images}
\label{LASCO}

\begin{figure}    
   \centerline{\includegraphics[width=\textwidth,clip=]{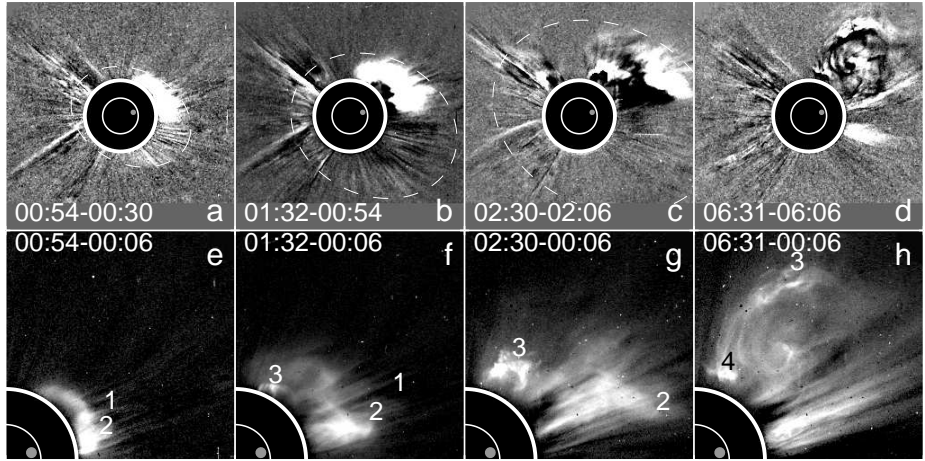}
              }
              \caption{
LASCO/C2 images of the related CME: running differences
(\textit{a--d}) and fixed-base differences (\textit{e--h}, NW
quadrant). All images are nonlinearly displayed. Thin circle: the
solar limb, thick circle: occulting disk, small gray disk: the
position of the eruption region, white dashed ovals outline the
expanding CME. Labels denote discussed features.               }
 \label{LASCO_fig}
   \end{figure}

LASCO/C2 \& C3 observed a CME with a central position angle of
$294^{\circ}$ in many frames from 00:54 onwards. The CME was
classified in the Preliminary 2004 SOHO LASCO Coronal Mass
Ejection List\footnote{ftp:/\negthinspace
/lasco6.nascom.nasa.gov/pub/lasco/status/LASCO\_CME\_List\_2004}
as a `partial halo'. Figure~\ref{LASCO_fig} presents four
LASCO/C2 images. A bright part of the CME has an angular span
$<90^{\circ}$. The spatial structure of the CME does not match a
classical three-component one. Its components do not resemble a
self-similar structure with each one being inside another one. No
cavity is detectable. The foremost part of the CME does not
resemble a well-defined frontal structure. It consists of diffuse,
faint material, inside which an interconnected structure is
visible (frame \textit{f}).

The fastest spike-like feature \textit{1} was used for
measurements in the SOHO LASCO CME
Catalog\footnote{http:/\negthinspace
/cdaw.gsfc.nasa.gov/CME\_list/}; its linear-fit speed is
409~km\thinspace s$^{-1}$. Labels \textit{2} and \textit{3} mark
the brightest components of the CME. Rotation of feature
\textit{3} determines its helical structure. The ejected material
is detectable around almost the whole occultor, but it is faint in
position angles $0-270^\circ$, even relative to the foremost NW
part (the fainter, faster SW part is listed in the SOHO LASCO CME
Catalog as another CME). White dashed ovals in
Figure~\ref{LASCO_fig} roughly outline the whole CME. Outflow of
the ejected material continues a few hours (\textit{e.g.}, feature
\textit{4} in frame \textit{h}).

Feature \textit{3} appears from behind the occultor at a position
angle corresponding to the trajectory of bright NoRH and TRACE
features. Their shapes are rather similar. The position angle of
feature \textit{2} roughly corresponds to TRACE feature
\textit{F2} and the bright sector visible in the EIT 195~\AA\
image at 00:24.

\subsection{Total Flux Radio Data}
\label{Radio_Data}

\begin{figure}    
   \centerline{\includegraphics[width=10cm,clip=]{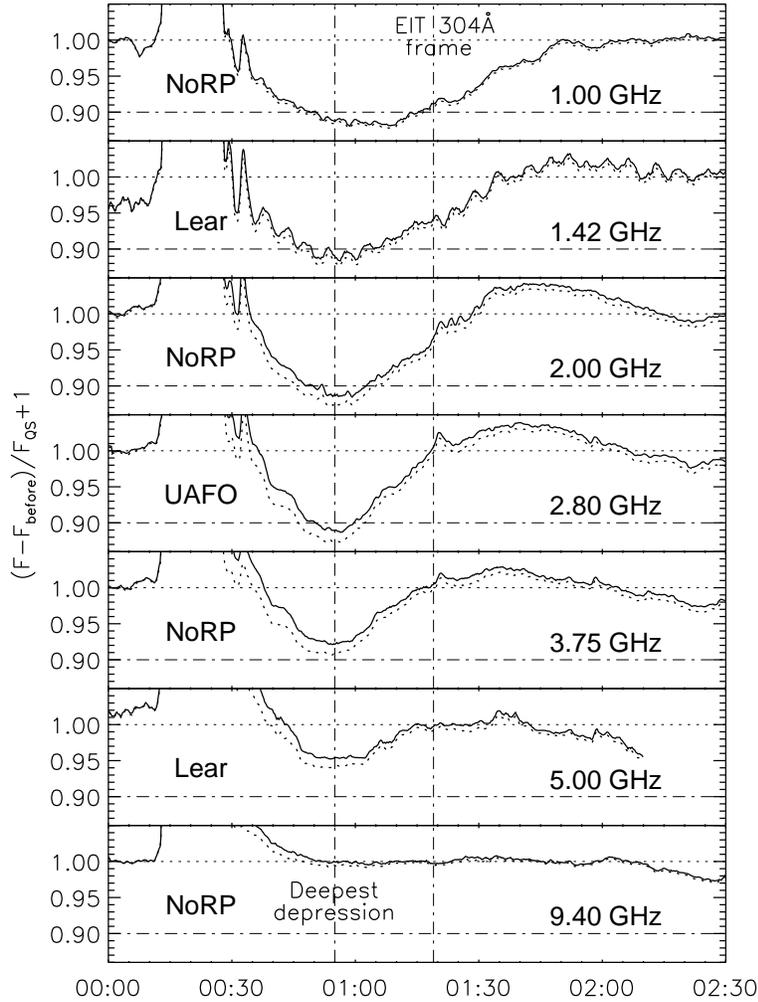}
              }
              \caption{
Total flux time profiles in a range of 1--9.4 GHz recorded in
Nobeyama, Learmonth, and Ussuriysk observatories. Dotted lines
show subtracted free-free emission of the arcade computed from
soft X-ray GOES data.               }
 \label{radio_timeprofs}
   \end{figure}

Records of total radio flux made with Nobeyama Radio Polarimeters
(NoRP) and in Learmonth and Ussuriysk Observatories show a strong
burst to start at about 00:12:20 and reaching 1200~sfu at 17 GHz.
Then, after 00:30--00:40, a post-burst decrease is observed at
frequencies $\leq 5$~GHz by 01:15--01:50
(Figure~\ref{radio_timeprofs}). The duration of the `negative
burst' is minimal at 5~GHz and increases toward lower frequencies
as well as the depression depth, from 5\% at 5~GHz to 12\% at
1~GHz. The deepest depression occurs at 00:55. The large-scale
darkening is registered in the EIT 304~\AA\ channel at the final
stage of the `negative burst'.

Thermal free-free emission of post-flare loops contributes to the
total flux. The dotted lines in Figure~\ref{radio_timeprofs} show
the total flux records without this contribution. To estimate it,
we computed emission measure and temperature, $T$, from soft X-ray
GOES data following White, Thomas, and Schwartz (2005). The size
of the soft X-ray emitting region was found from GOES/SXI images,
and a geometrical depth, $L \simeq A^{1/2} \approx 1.3 \cdot
10^9$~cm, estimated from its area, $A$. The optical thickness at a
frequency $\nu$, $\tau_\nu = k_\nu L$, was determined from an
expression
\begin{eqnarray}
\tau_\nu \approx 0.2n_e^2 \nu^{-2} T^{-3/2} L \quad \mathrm{(CGS\
units).} \label{optical_thickness_eqn}
\end{eqnarray}

\subsection{Overall Picture of Near-Surface Activities}
\label{Overall_Picture}

\begin{figure}    
   \centerline{\includegraphics[width=10.cm,clip=]{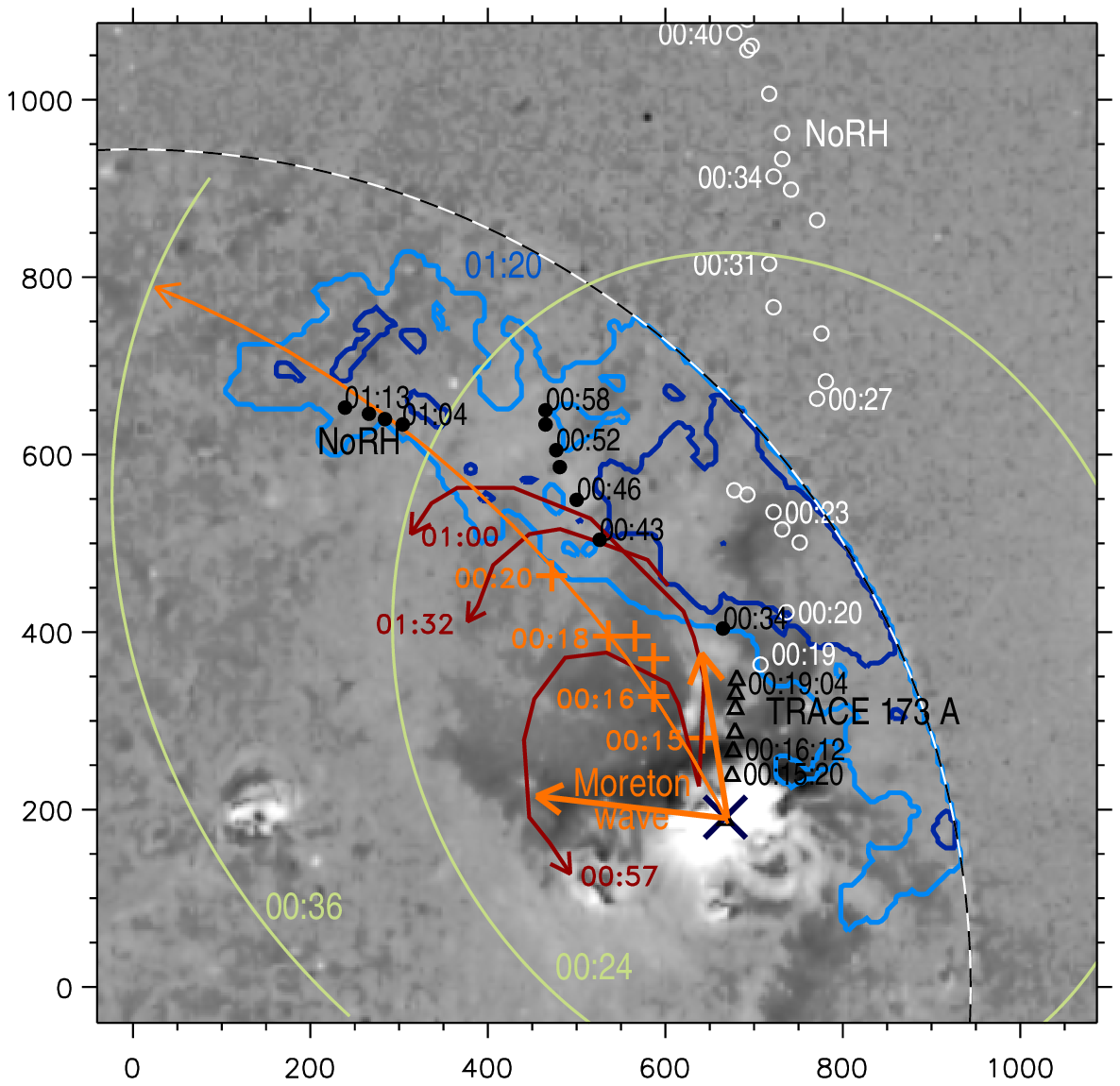}
              }
              \caption{
An overall map of large-scale disturbances observed during the
event on the solar disk and in its vicinity. Gray scale
background: the EIT 195~\AA\ difference image (00:24--00:00). Blue
contours: darkening in the EIT 304~\AA\ image (13 Jul, 01:20 $-$
12 Jul, 19:20); bright $-20\%$, dark $-60\%$. Green-yellow oval
and arc: wave fronts in the EIT 195~\AA\ difference images (00:24
and 00:36). Black slanted cross: eruption center. Red crosses:
measured positions of the Moreton wave fronts. Red arc with the
arrow: approximate propagation direction of the wave along the
solar surface; red arrows roughly show its angular span. Three
brown arrows: trajectories of moving fragments, which were
revealed from MLSO H$\alpha$ images. Triangles: the leading edge
of the expanding bright feature in the TRACE 173~\AA\ images.
Circles: measured positions of the moving bright (white) and dark
(black) features in NoRH 17 GHz images. Black-white dashed circle:
the solar limb.               }
 \label{Scheme}
   \end{figure}

Figure~\ref{Scheme} presents an overall picture of large-scale
disturbances observed on the solar disk and in its closest
vicinity during and after the 13 July 2004 eruptive event. The
figure summarizes observations in the H$\alpha$ line, in 195~\AA,
304~\AA\ (EIT), and 173~\AA\ (TRACE) bands as well as at 17 GHz
(NoRH).

The event starts at 00:02:30 with stretch of a filament or two
`merged' ones. During its steady rise, flare energy release starts
by 00:13:40 beneath. The filament explosively erupts at about
00:15. The explosion produces Moreton and `EIT' waves, which run
faster than all other observed disturbances. Dark fragments of the
filament form surges or fly mostly north, even reaching vicinities
of the North Pole. One portion of the bright material moves up in
the plane of sky, while another one moves west-northwest. The
former portion is traceable in TRACE and NoRH images from the
eruption site up to the edge of the field of view. The latter
portion is only detectable in EIT 195~\AA\ images.

\section{Absorption}
\label{Absorption}

As multi-spectral data show (Figure~\ref{Scheme}), fragments of
the eruptive filament dispersed over a large area could be
responsible for absorption phenomena observed in radio range
(`negative burst'), H$\alpha$ (surges), and EUV at 195~\AA\ and
304~\AA. In this section, we estimate parameters of the absorbing
material and find out if it is possible to reconcile all mentioned
manifestations. We then compare the estimated masses of the
absorber and the CME.

\subsection{Simulation of Radio Absorption}
\label{Simulations}

  \begin{figure}    
   \centerline{\includegraphics[width=10.cm,clip=]{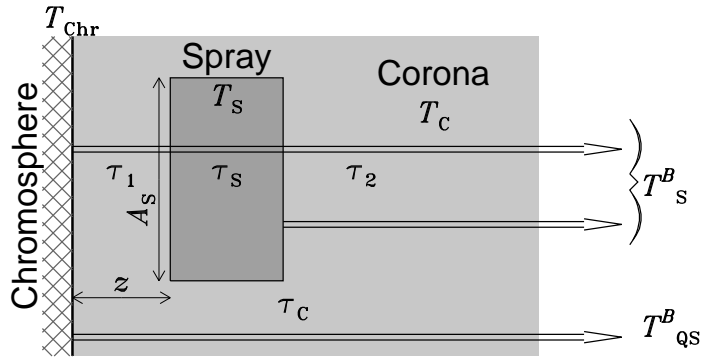}
              }
              \caption{
A simple model to estimate the frequency dependence of radio
absorption.               }
 \label{Model_fig}
   \end{figure}

We simulate the `negative burst' using a simple four-layer model
with a uniform absorbing slab (Figure~\ref{Model_fig}). The model
contains 1)~the chromoshpere, 2)~the `spray' of area $A_S$
`inserted' into the corona, 3)~a coronal layer of depth $z$ between
the chromosphere and spray, and 4)~a coronal layer between the spray
and observer. The ratio of the observed flux to the quiet Sun's
total flux is
\begin{eqnarray}
F/F_\mathrm{QS} =
[T^{B}_\mathrm{QS}(A-A_\mathrm{S})+T^B_\mathrm{S}
A_\mathrm{S}]/(T^{B}_\mathrm{QS}A). \label{flux_ratio_eqn}
\end{eqnarray}
\noindent $A(\nu)$ is the quiet Sun's area. We used frequency
dependencies of the quiet Sun's radio radius and brightness
temperature found by Borovik (1994). Superscripts `$B$' denote
brightness temperatures to distinguish them from kinetic ones.
Since the brightness temperature after each layer is a sum of its
own emission and a non-absorbed remainder emission from preceding
layers, the observed brightness temperature, $T^B_\mathrm{S}$, of
the spray with a kinetic temperature $T_\mathrm{S}$ is
\begin{eqnarray} T^B_\mathrm{S} = T_\mathrm{Chr}e^{-(\tau_1 + \tau_2 +
\tau_\mathrm{S})}
+ T_\mathrm{C}(1-e^{-\tau_1})e^{-(\tau_2+\tau_\mathrm{S})} \label{model_eqn}\\
+\ T_\mathrm{S}(1-e^{-\tau_\mathrm{S}})e^{-\tau_2} +
T_\mathrm{C}(1-e^{-\tau_2}), \nonumber
\end{eqnarray}
\noindent where $\tau_i$ is determined by
expression~(\ref{optical_thickness_eqn}), $\tau_1 = \tau_\mathrm{C}
- \tau_2$, with $\tau_\mathrm{C}$ being the total optical thickness
of the corona at a given frequency, $\tau_2 = \tau_\mathrm{C}
e^{-2z/H}$, $H = 2kT_C/(m_i g_\odot) \approx 8.4 \cdot 10^9\,
\mathrm{cm}$ the height of the uniform atmosphere, $k$ Boltzmann
constant, $z$ the height of the absorbing layer above the
chromosphere, $m_i$ the average mass of ions, and $g_\odot =
GM_\odot/R^2_\odot$ the gravity acceleration at the photosphere, $G$
the gravity constant, $M_\odot$ the mass of the Sun.

\begin{figure}    
   \centerline{\includegraphics[width=8.cm,clip=]{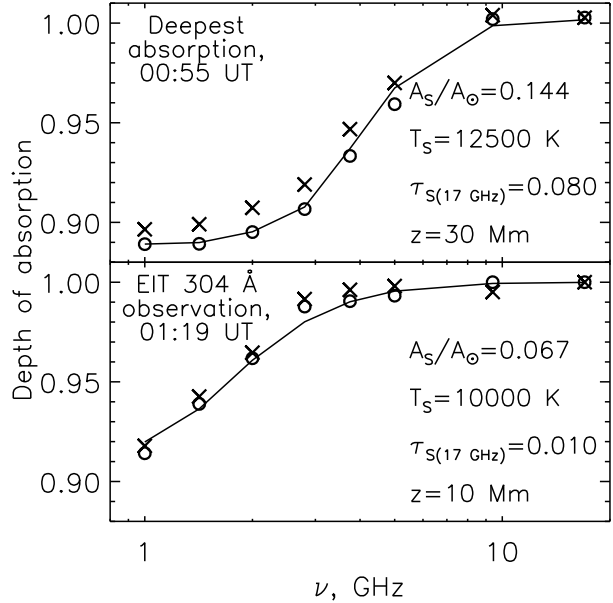}
              }
              \caption{
Radio absorption measured directly (crosses) and with subtracted
contribution from the arcade (circles) along with results of
modeling (lines) for two instants.               }
 \label{absorp_spec}
   \end{figure}

Using this model, we fitted the radio absorption depth measured
from multi-frequency records (Figure~\ref{radio_timeprofs}) for
two instances, the maximum radio absorption (00:55) and EIT
observation at 304~\AA\ (01:19). The results are shown in
Figure~\ref{absorp_spec} with best-fit parameters. The model is
rough; nevertheless, variations of parameters of $\pm 20\%$ cause
detectable discrepancies with observations.

With an optical thickness evaluated, a density and mass of the
absorber could be estimated if its geometrical depth was known.
The latter can be found from several facts. One estimate is the
POS distance of the overturn of feature 3a, 190~Mm
(Section~\ref{H_alpha}); with the position of the active region,
it is expected to be comparable with its height. The time of the
overturn, 00:38--00:40, provides a better estimate. Since the
maximum height is large, we used general expressions with an
inconstant $ g(R) = GM_\odot/R^2$ for a free fall time from the
heliocentric distance $r$ to $R$, velocity $v$ at $R$, and $r(v)$:
 \begin{eqnarray}
t =  \sqrt{\frac r{2g(R)R^2}}\left\{ \sqrt{R(r-R)}+ r/2\left[\pi/2
+\arcsin(1-2R/r)\right] \right\},  \label{free_fall} \\
 v = \sqrt{2g(R)R(1-R/r)}, \quad r = 2g(R)R^2/[v^2 + 2g(R)R]. \nonumber
 \end{eqnarray}
Using these expressions, we find the height of the overturn to be
180\,--\,200~Mm and the initial velocity of
350\,--\,380~km\thinspace s$^{-1}$, reasonably close to the measured
POS parameters of the filament. From the height and time of the
overturn, a height at 00:55 is 100\,--\,130~Mm. With $z \simeq
30$~Mm estimated from the model, the geometrical depth is
70\,--\,100~Mm, and from expression~(\ref{optical_thickness_eqn}) we
find the density of $(1.1-1.3)\cdot 10^8$~cm$^{-3}$ and the mass of
$(3-4)\cdot 10^{15}$~g at that time.

It is difficult to estimate the depth at 01:19 from observations.
We speculate that the height at that time was significantly less
than it at 00:55 was --- \textit{e.g.}, $<50$~Mm. On the other
hand, there is no reason for the absorbing cloud to be a thin film
located at a height of 10~Mm (from the model); therefore, we
assume its depth to be $> 2$~Mm (most probable 10\,--\,20~Mm).
With these depths, we get limits for the density of
$(0.5-2.3)\cdot 10^8$~cm$^{-3}$ and mass of $(0.8-3.5)\cdot
10^{14}$~g; most probable are $\simeq 10^8$~cm$^{-3}$ and $\simeq
2\cdot 10^{14}$~g.

\subsection{Absorption in H$\alpha$ and EUV}

H$\alpha$ opacity is due to neutral hydrogen absorbing continuum
radiation of underlying atmosphere. Mein \textit{et al.} (1996);
Heinzel, Mein, and Mein (1999); Molowny-Horas \textit{et al.}
(1999), and Tziotziou \textit{et al.} (2001) calculated physical
parameters in stationary or slowly moving cool clouds like
filaments on the basis of a non-LTE radiative transfer approach.
Heinzel, Schmieder, and Tziotziou (2001) and Heinzel \textit{et
al.} (2003a) studied relations between opacities in H$\alpha$ and
Lyman continuum at 912~\AA\ and proposed an approximate expression
to estimate the hydrogen density from the optical thickness at the
H$\alpha$ line center. The Doppler-brightening effect decreases
absorption of moving prominences, when line-of-sight velocities
exceed $\sim 50$~km\thinspace s$^{-1}$. The Doppler shift of the
H$\alpha$ absorption line exceeds the bandwidth of the BBSO
H$\alpha$ 0.5~\AA\ Lyot filter at a lower velocity of $\sim
20$~km\thinspace s$^{-1}$.

A spectroscopic model for absorption of EUV emission in filaments
and prominences was developed by Heinzel \textit{et al.}
(2003a,b), Anzer and Heinzel (2005), and Schwartz \textit{et al.}
(2006). They showed that the EUV opacity is due to photoionization
of hydrogen (below the edge of the Lyman continuum, 912~\AA),
neutral helium ($<504$~\AA), and ionized helium ($<228$~\AA).
Absorption of continuum does not depend on radial or turbulent
velocity. For 195~\AA, we take an average absorption cross-section
of a `cloud' consisting of 92\% of H and 8\% of He, $\sigma_{195}
\approx 7 \cdot 10^{-20}$~cm$^2$, which varies within 30--40\% in
a temperature range of $0 < T < 80\,000$~K (we calculated the
absorption cross sections for 195 and 304~\AA\ using Table~2 and
expressions (7) and (9) from Anzer and Heinzel, 2005). Along with
absorption, darkening in coronal lines could be due to a volume
blocking effect (VBE), \textit{i.e.}, the absence of a coronal
radiation from the volume of the cloud. This effect depends on the
height and depth of the cloud relative to a height scale at a
given wavelength. For 195~\AA, a typical radial height scale is
$\sim 70\,$Mm, and for our position angle, $42^{\circ}$, the
line-of-sight height is $H_\parallel \sim 100\,$Mm.

\begin{figure}    
   \centerline{\includegraphics[width=10.cm,clip=]{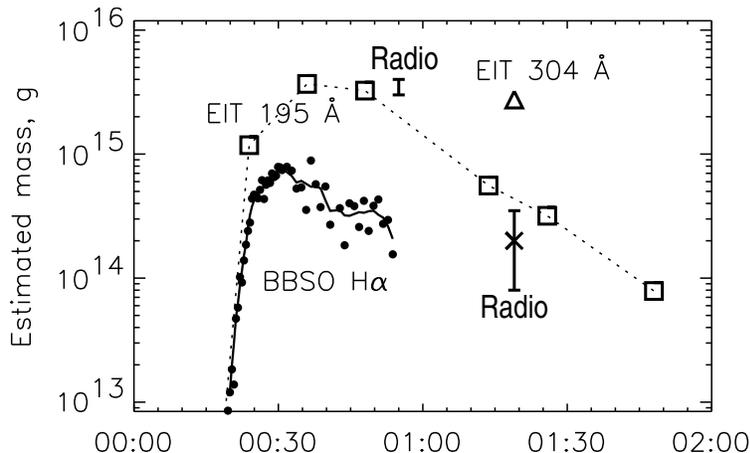}
              }
              \caption{
The mass of the ejecta estimated from opacities in H$\alpha$
(circles), 195~\AA\ (squares), 304~\AA\ (triangle), and radio
range (bars: limits, cross: probable).                }
 \label{ha_195_mass}
   \end{figure}

We computed the hydrogen density from the optical thickness in the
H$\alpha$ line center assuming a line-of sight velocity
$V_{\parallel} < 20$~km\thinspace s$^{-1}$, a turbulent one
$V_\mathrm{turb} = 5$~km\thinspace s$^{-1}$, temperature 8\,000~K,
and depths of absorbing fragments to be equal to their widths.
Figure~\ref{ha_195_mass} presents masses estimated from absorption
of all emissions considered. The mass estimated from H$\alpha$
increases from 00:14 by 00:30 and then decreases likely because
velocities of fragments become $V_\parallel>20$~km\thinspace
s$^{-1}$.

To estimate the opacity at 195~\AA, we analyzed EIT difference
images (subtrahend is 12 July, 23:59:57) in the interval of
00:12--02:00. Regions of probable absorption were found as
depressions below $1\sigma$ level, where $\sigma$ was computed as
an r.m.s. over negative pixels of darkening regions in the NW
quadrant only. The regions revealed in this way contained
long-lived dimmings (\textit{e.g.}, \textit{D1, D2}), which were
unlikely to be due to absorption. We identified them from the
difference image at 02:00 and excluded from computations. The mass
estimated from 195~\AA\ images reaches a maximum of $3.7 \cdot
10^{15}$~g at 00:35 and then diminishes. The estimations miss some
fragments of the ejecta that crossed the long-lived dimmings and
could contain 20--30\% of the total mass. A noticeable part of the
ejected mass could be contained in hot ionized plasmas of the
transition layer between cold absorbing material and ambient
corona (analogous to an EUV filament channel) seen in TRACE
173~\AA\ as a bright contour. We did not consider the VBE, because
the depth of absorbing fragments was $\ll H_\parallel$.

A large depression area at 304~\AA\ (01:19, see
Figure~\ref{EIT_304}) shows a part of ejected material to return
back onto the Sun. The same selection criteria as for 195~\AA\
were applied. A mean brightness ratio to the pre-event image in
the whole region is about 0.54, while in its significant part the
ratio is noticeably below 0.5.

Absorption at 304~\AA\ can be due to photoionization of H and
He{\sc i} by continuum or due to resonant scattering of the He{\sc
ii} line in He ions. The latter mechanism cannot provide the
absorption ratio $< 0.5$: the cloud emits resonant photons in
$4\pi$ angle, while incident photons come from below and lateral
directions, \textit{i.e.}, $2\pi$. Also, the resonant scattering
is very sensitive to the line-of-sight velocity: the Doppler shift
exceeds a typical width of the He{\sc ii} 304~\AA\ line if
$V_\parallel > 50$~km\thinspace s$^{-1}$. Absorption of the
continuum is more probable (cross section $\sigma_{304} \approx
5.5 \cdot 10^{-19}$~cm$^2$), and the mass is $\sim 2.8 \cdot
10^{15}$~g (triangle in Figure~\ref{ha_195_mass}).

The closest 171~\AA\ (01:00) and 195~\AA\ (01:25) images do not
show absorption in the 304~\AA\ darkening region (absorption at
195~\AA\ occurs in other regions). At this time, the cloud was
likely in between of the heights of He{\sc ii} 304~\AA\
illumination and coronal loops responsible for the 171 and
195~\AA\ emissions, as absorbing flows in TRACE 173~\AA\ images
show. Low clouds with a temperature close to the chromospheric
one, 10\,000 K, are not detectable in radio range also.

The CME mass estimated using LASCO software reaches $1.3 \cdot
10^{15}$~g at 04:06. Correction due to the off-plane direction of
the CME (Vourlidas \textit{et al.}, 2000) of $\sim 1.4$ gives its
real mass of $1.8 \cdot 10^{15}$~g. A lesser mass of $1.8 \cdot
10^{14}$~g in the SOHO LASCO CME Catalog is related to the faint
SW part (see Figure~\ref{LASCO_fig}a--c).

The estimates of the mass of the ejecta from absorption of
different emissions agree with each other within their accuracy
and limitations of methods. The masses of the CME and the returned
part are close. Thus, the ejecta disintegrated into parts with
comparable masses, one of which flew away, while another one fell
back onto the Sun. One might ask if the absorbing cloud left the
Sun later and was observed as feature \textit{4} in
Figure~\ref{LASCO_fig}h. However, its estimated mass is $1.1 \cdot
10^{14}$~g, much less than the mass of the cloud. This rules out
such a possibility.

\section{Kinematics}
\label{Kinematics}

In this section, we choose analytic expressions to describe
kinematics of a wave expanding from the eruption site and mass
fragments flying away. Their birth in a single process is certain,
but their subsequent interrelation is vague, \textit{i.e.}, it is
not clear if a piston existed to determine the motion of the wave.
The oval shape of the wave front disfavors the role of the faint
NW sector in the EIT image at 00:24 as a piston. The wave and
fragments appear to move independently. Deceleration of mass
fragments, constituting the CME, points at importance of gravity.
This also shows up in the fact that a part of the ejected mass
falls back and produces a cool absorbing `screen' to cover a
significant part of the Sun.

\subsection{Approach}

We use a self-similar approximation of the expansion of mass
fragments. Low (1982, 1984a,b) first applied it to expansion of a
CME. Uralov, Grechnev, and Hudson (2005) derived expressions
convenient to compare with observations that apply independent of
the presence of a CME-driven coronal shock. If it is present, then
CME parts are assumed to be confined by a contact surface, which
is also a surface of a moving piston. A solution belongs to a
class of self-similar motions, in which Lagrangian velocities of
pieces of mass $v$ are proportional to their distances $r$ from
the center of expansion, $v \propto r$ (see also Sedov, 1981, p.
318).

To describe wave-like phenomena, we employ concepts of
self-similar theory of a point-like explosion (\textit{e.g.},
Sedov, 1981; Zel'dovich and Raizer, 1966). Such an explosion
produces a strong shock wave in the ambient medium.

A crude example to combine both these approaches is an air
explosion of a bomb consisting from a charge and a shell. The
outcome is a shock wave and fragments of the shell flying away.
They expand synchronously during the explosion and move
independently afterwards.

\subsection{Mass Fragments}

\begin{figure}    
   \centerline{\includegraphics[width=10.cm,clip=]{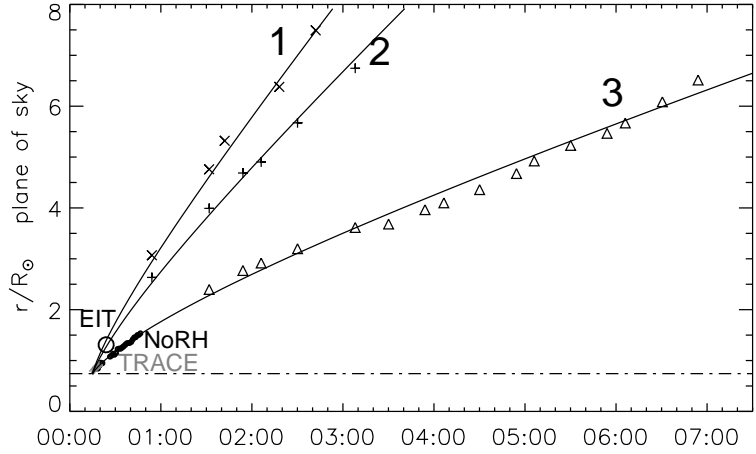}
              }
              \caption{
Distance-time plots (in solar radii) of fragments observed with
TRACE at 173~\AA\ (gray crosses), NoRH at 17 GHz (circles), and
LASCO/C2 (black; slanted crosses for feature 1, straight crosses
for feature 2, and triangles for feature 3). The large circle
denotes the observation point of the bright aureole in the EIT
195~\AA\ image at 00:24. Dash-doted line denotes the position of
the eruption center.                }
 \label{frag_plots}
   \end{figure}

As Section~\ref{Overall_Picture} shows, some counterparts of CME
features are detectable on the Sun or nearby.
Figure~\ref{frag_plots} shows distance-time points measured in POS
from TRACE, NoRH, and LASCO data. The best coverage has feature 3
(see Section~\ref{LASCO}) and its counterparts. A quasi-periodical
component in plots of all LASCO features is due to their rotation.
It complicates measurements and decreases their accuracy.

Expression (A8), $v^2 = v_0^2+2A/r_0\left({1- r_0/r}\right)$, from
Uralov, Grechnev, and Hudson (2005) with $r \to \infty, v \to
v_\infty$ and $v_0 = v(r = r_0)$ gives $2A/r_0 = v_\infty^2 -
v_0^2$:
\begin{eqnarray}
v^2 = v_0^2+\left(v_\infty^2 - v_0^2\right)\left({1-r_0/r}\right).
\label{fragments_expansion}
\end{eqnarray}
This expression shows that with the increasing distance, the
deceleration of all features monotonically decreases up to zero.
Therefore, no polynomial fit of the measured data points applies.
To compare results with observational data, a few parameters are
necessary; the initial and asymptotical velocities, $v_0$ and
$v_\infty$; the initial size of the ejecta, $r_0$; and origins of
measurements both in space and time. The origins are known with a
satisfactory accuracy (see Section~\ref{Shock_wave}; start time of
00:14:50). The $v_\infty$ can be estimated from
Figure~\ref{frag_plots} (it is 409~km\thinspace s$^{-1}$ for
feature 1 according to the SOHO LASCO CME Catalog). The $v_0$ for
feature 3 can be found from TRACE and NoRH data points. Obviously,
$v_0^{(1)} > v_0^{(2)} > v_0^{(3)}$. The $r_0$ parameters could be
only found in attempts to obtain a good fit. The curves in
Figure~\ref{frag_plots} were obtained with the following
parameters:

LASCO feature 1: $r_0 = 270$~Mm, $v_0 = 1100$~km\thinspace
s$^{-1}$, $v_\infty = 409$~km\thinspace s$^{-1}$;

LASCO feature 2: $r_0 = 220$~Mm, $v_0 = 850$~km\thinspace
s$^{-1}$, $v_\infty = 310$~km\thinspace s$^{-1}$;

LASCO feature 3: $r_0 = 110$~Mm, $v_0 = 490$~km\thinspace
s$^{-1}$, $v_\infty = 100$~km\thinspace s$^{-1}$.

These parameters provided a satisfactory fit for each of the three
features visible in LASCO images, with the accuracy being worse
for features 1 and 2, whose counterparts are poorly identified
at/near the Sun.

\subsection{Wave -- Uchida's Model: Weak Fast Mode}

The almost perfect oval shape of the brightening in the EIT
195~\AA\ image at 00:24 and its expansion visible in the next
image at 00:36 suggest its wave-like origin and association with
the Moreton wave. The following facts also support this: (i)~its
configuration is different from the main CME components, in
particular, (ii)~its eastern part is roughly as bright as the
western part, if not exceeds it, whereas the eastern CME part in
LASCO/C2 images is much fainter than its western part, and
(iii)~the apparent velocity of the oval, $\simeq 700$~km\thinspace
s$^{-1}$, is comparable with the Moreton wave's one. This
possibility was previously discussed by, \textit{e.g.}, Hudson and
Warmuth (2004) and Warmuth \textit{et al.} (2001, 2004a, 2004b).

According to the original Uchida's (1968) idea, Moreton waves are
excited by coronal waves, being their skirts. Uchida (1968)
considered a weak short fast-mode MHD wave to propagate in a
radially-symmetric corona, with the Alfv{\' e}n speed increasing
with height. This caused refraction of rays into regions of lower
Alfv{\' e}n velocity, toward the solar surface, and turned the
wave front down. This model predicted acceleration of Moreton
waves. However, Warmuth \textit{et al.} (2001, 2004a) found their
systematic deceleration (Yamaguchi \textit{et al.} (2003) reported
acceleration of a Moreton wave in its propagation into a coronal
hole). Their observed velocities are sometimes too high for weak
fast-mode MHD waves. These inconsistencies with observations do
not appear to rule out weak fast-mode MHD waves completely, but
encourage to search for another possibility to match observations
better. Warmuth \textit{et al.} (2001, 2004a, 2004b) and Hudson
and Warmuth (2004) proposed blast waves to be responsible for
Moreton waves and, probably, some `EIT waves'.

\subsection{Wave -- Strong Shock}
\label{Shock_wave}

We use estimative expressions from a theory of a strong point-like
explosion in a variable-density medium. A rigorous analysis of this
gas dynamic task was carried out by Sedov (1981). Self-similarity of
the solution is ensured by a large pressure excess inside the volume
confined by the shock front over non-disturbed medium. If the
magnetic field in ambient plasma is strong, then the profile and
shape of the shock front change. This change is small for a strong
shock, because the increase of the magnetic pressure in the shock is
determined by extreme plasma compression, whereas gas pressure is
unlimited. Also, the magnetic field in our case is weak, because the
disturbance propagates over quiet Sun's regions.

Kinematics of a single-pulse spherical blast wave is governed
mainly by pile-up mass. We consider propagation of a self-similar
blast wave excited by an explosion of an energy $E$ in media (a)
with constant density, and (b) with a radial density falloff from
the explosion center, $\rho \propto r^{-\alpha}$.

\begin{enumerate}
\item
 For a constant density, $\rho = \rho_0$: $\rho_0 R^3 V^2 =
const\ E$ ($R$ radius and $V$ velocity of the shock front), with
$const$ corresponding to constancy of potential to kinetic energy
ratio for plasma involved into self-similar motion. Thus, $V =
\frac{dR}{dt} = \left( \frac{const\ E}{\rho_0 R^3}\right)^{1/2}
\propto R^{-3/2}$, and $R \propto t^{2/5}$.

\item
 Similarly, for $\rho = b r^{-\alpha}$:
$V^2\int\frac{b}{r^\alpha}r^2dr \to V^2 R^{[3-\alpha]} = const\
E$. Consequently, $V = \frac{dR}{dt} \propto R^{-[(3-\alpha)/2]}$,
and
\begin{eqnarray}
R \propto t^{[2/(5-\alpha)]} =t^\delta.
\label{wave_expansion}
\end{eqnarray}
\end{enumerate}
 A strong spherical shock decelerates if $\alpha <3$
and accelerates if $\alpha >3$. Thus, $\delta = 2/(5-\alpha)$,
with $\delta = 0.4$ for a homogeneous medium ($\alpha =0$) and
$\delta = 0.67$ for a medium with $\rho \propto r^{-2}$. To
compare expression~(\ref{wave_expansion}) with measured points of
the Moreton wave, we use expression~(\ref{wave_expansion}) in a
form $(R-R_0)/(R-R_1) =
\left[(t-t_0)/(t-t_1)\right]^{[2/(5-\alpha)]}$. Here $R_0$ is the
distance between the origin of measurements and explosion center,
$t_0$ the corresponding time, and $t_1$, $R_1$ are related to one
of measured points.

\begin{figure}    
   \centerline{\includegraphics[width=10.cm,clip=]{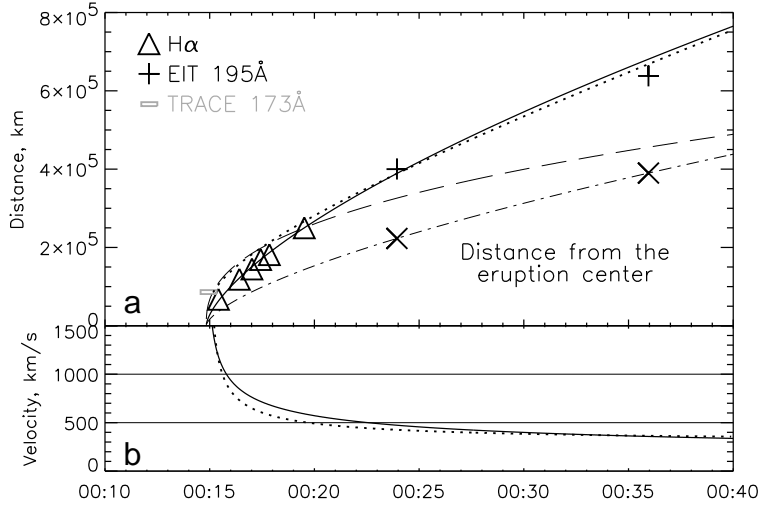}
              }
              \caption{
(a)~POS Distance-time plots of the Moreton wave (triangles) and
EIT oval (crosses) measured in the same direction from the
eruption center. Rectangle denotes the first manifestation of a
wave in a TRACE 173~\AA\ image. Curves: strong shock with density
distributions of $\rho = const$ (dashed) and $\rho \propto r^{-2}$
(solid); weak shock with $\rho = const$ (dotted). Slanted crosses:
distances of the EIT oval from the eruption center measured toward
the solar disk center; dash-dotted curve shows their fit with
$\rho \propto r^{-2}$. (b)~Instantaneous velocities computed for a
strong shock with $\rho \propto r^{-2}$ (solid) and a weak shock
with $\rho = const$ (dotted).}
 \label{fig_wave1}
   \end{figure}

Figure~\ref{fig_wave1} shows the results. Triangles denote the
Moreton wave, straight crosses denote distances of the EIT oval
from the eruption center measured in the same direction, slanted
crosses show measurements of the EIT oval toward the solar disk
center, and a rectangle presents the first manifestation of the
wave in a TRACE 173~\AA\ image (width: exposure time, height:
positional uncertainty). The Moreton wave is expected to slightly
lag behind the coronal wave (Vr{\v s}nak \textit{et al.}, 2002),
as the TRACE and H$\alpha$ fronts show. We used the last point of
the Moreton wave as a reference one $(t_1, R_1)$. The dashed curve
corresponding to $\alpha = 0$ shows much stronger deceleration
than actually observed. A good fit is achieved with $\alpha = 2$
($\delta = 0.67$) and a start time of 00:14:50$\pm 20$~s for both
the Moreton and `EIT' waves. The positional uncertainty of the
explosion center is $\lsim 20$~Mm; origins of both the position
and time are close to the observed eruption ones. \textit{The
difference between distances measured along the arc on the sphere
in Figure~\ref{Scheme} and POS distances in the same direction
does not exceed measurement errors.}

The kinematical closeness of the Moreton and `EIT' waves confirms
their common nature, and agreement with the calculated plot
supports their origin due to coronal blast shock. Note that data
points of the `EIT wave' suggest a stronger deceleration, implying
a density falloff $\alpha < 2$ at longer distances.

The calculated instant shock speed in Figure~\ref{fig_wave1}b can
be compared with estimates from type II bursts listed in Solar
Geophysical Data\footnote{http:/\negthinspace
/sgd.ngdc.noaa.gov/sgd/jsp/solarindex.jsp}. The estimates decrease
from 1000~km\thinspace s$^{-1}$ at 00:17--00:20 (Culgoora) via
850~km\thinspace s$^{-1}$ (00:16--00:40, Learmonth) to
550~km\thinspace s$^{-1}$ at 00:27--00:41 (Holloman). Such
estimations use model plasma densities, which are uncertain.
Nevertheless, the calculated shock speed and estimations from type
II bursts are in overall agreement. This supports the same origin
of the Moreton wave, `EIT wave', and the type II radio burst,
likely associated with the same decelerating coronal blast shock,
as Warmuth \textit{et al.} (2001, 2004a, 2004b) concluded
(however, Pohjolainen, Hori, and Sakurai (2008) interpret this
type II burst to be due to two distinct shocks).

Our plot in Figure~\ref{fig_wave1} was obtained under assumptions
of (1)~a strong shock and (2)~an omnidirectional density falloff
from the eruption site. Their agreement with observations appears
to be surprising, because a horizontal density falloff is not
expectable at large distances. We discuss this issue in the next
section.

We also fitted the oval envelope of the CME with $\rho \propto
r^{-2.7}$ expectable for a coronal shock propagating outward: at
heights of $(0.2-10)R_\odot$, the density in the corona above an
active region falls off $\propto r^{-2}$ ($r$ the distance from
the photosphere) according to Newkirk model (Newkirk, 1961), and
$\rho \propto r^{-2.9}$ from a model compiled by Gary (2001). This
fit is close to the motion of LASCO feature~\textit{1}. Note that
with $v_\infty^2 \ll  v_0^2$ and $\alpha = 2$,
expressions~(\ref{fragments_expansion}) and (\ref{wave_expansion})
predict the same height-time plots. Poor measurement accuracy does
not allow to distinguish between the wave and mass fragments.
Their closeness is also possible.

\subsection{Comments on a Realistic Situation}

We come to the following possibilities for the Moreton wave.

1. \textit{Uchida's (1968) refraction model} is a linear acoustic
approximation, which is very sensitive to variations of plasma
parameters. This model predicts acceleration of a wave that does
not correspond to the observed situation.

2. \textit{Strong shock wave}: gas velocity behind the wave front
(Lagrangian velocity; it is presented by subsequent motion of loop
$L1$ in TRACE images, Figure~\ref{TRACE}, $350$~km\thinspace
s$^{-1}$) is much higher than the ambient fast magnetosound speed,
$v_\mathrm{f}$. This strongly non-linear wave is mainly sensitive
to density distribution. Deceleration naturally appears and
critically differs if the wave runs along, against, or across the
density gradient. To explain propagation of the Moreton wave, we
assumed a spherically-symmetric coronal density falloff from the
explosion center. In this case, the wave runs contrarily to the
density gradient, and its deceleration is determined by expression
(\ref{wave_expansion}). This resulted in a good fit with $\alpha =
2$. Otherwise, the deceleration is too strong, as the following
considerations show.

The Moreton wave runs horizontally, presumably being a skirt of a
coronal shock front. With a constant density at a fixed height, a
strong shock decelerates even faster than
expression~(\ref{wave_expansion}) with $\alpha =0$ predicts. To
roughly estimate this case, we consider a vertically-stratified
atmosphere with an upward density falloff of $\rho \propto
r^{-\alpha}$. Here, the coronal part of the Moreton wave runs
across the density gradient. The front of a blast wave in such
atmosphere is not a spherical one; its initially vertical parts in
propagation progressively incline down at the same height. We
assume pressure $P$ to be uniform inside a volume confined by the
shock front, $P \propto \frac{ E}{ R^3}$, with $ R$ being the
vertical extension of the wave surface. Its variation is
determined by expression~(\ref{wave_expansion}). In a strong
shock, $\rho_x V_x^2 \simeq P$, where $ V_x = \frac{dR_x}{dt} $ is
a velocity of the front along a constant-density surface, $\rho_x
= const$. With $ \alpha =2$, $ V_x  \propto t^{-1} $, and $ R_x
\propto \ln(t)$. After the fast onset, such a shock rapidly
decelerates up to $v_\mathrm{f}$ (cf. a typical speed of `EIT
waves').

Besides a spherically-symmetric density falloff, a high horizontal
deceleration of a strong shock could be prevented if a piston
moved properly in the same direction (cf. Low, 1984a). If a
self-similar solution does exist, then a piston and wave of a
common origin diverge slowly. Indeed, the velocity of the strong
shock front is $ \frac{dR}{dt} = \frac{\gamma +1}{2}
v_\mathrm{sh}$ with $\gamma $ the polytropic index and
$v_\mathrm{sh}$ the Lagrangian plasma velocity just behind the
shock front. The velocity of the contact discontinuity is $
\frac{dr_\mathrm{c}}{dt} = v_\mathrm{c}$ with $v_\mathrm{c}$ the
Lagrangian plasma velocity within the discontinuity. In such
self-similar motions, $ \frac{ v_\mathrm{sh} }{ v_\mathrm{c} } =
\frac{R}{r_\mathrm{c}} $. From these expressions, $ \frac{dR}{R} =
\frac{\gamma +1}{2} \frac{dr_\mathrm{c}}{r_\mathrm{c}}$, and the
ratio of distances of the shock front $R$ and the piston $r_c$
from the explosion center is $ \frac{R}{r_\mathrm{c}} \propto
r_\mathrm{c}^{(\gamma -1)/2} $. Initially $R= r_c=1$, and then the
ratio increases slowly; both the piston and wave are expectable to
be not far from each other in images, which we showed. However,
the wave appears to have run far away well before the first
appearance of mass behind it. A blast wave is also favored because
the initial velocities of the ejecta do not point in the same
direction as the EIT oval.

An omnidirectional density falloff seems to be possible near the
active region, but unlikely at long distances. A possible solution
of the problem is subsequent damping of a strong shock to a
moderate or weak one. The weak shock runs along rays determined by
the spatial distribution of $v_\mathrm{f}$, which, in turn,
depends on the density distribution. The speed of the weak shock
depends on the density gradient along the ray (\textit{e.g.},
horizontal) rather than on the crosswise (vertical) gradient. Just
the vertical density gradient caused the extreme deceleration of a
strong shock propagating horizontally, that was considered above.

3. \textit{Weak shock}: gas velocity behind the wave front is
$<v_\mathrm{f}$. Propagation of a spherical weak shock in a
uniform plasma calculated using expressions from a paper of
Uralova and Uralov (1994) is shown in Figure~\ref{fig_wave1} by
dotted lines for $v_\mathrm{f} = 300$~km\thinspace s$^{-1}$ and
the phase speed of the wave at its first manifestation of
$1600$~km\thinspace s$^{-1}$. This curve matches the coronal wave,
whereas at shorter distances its agreement with the Moreton wave
is worse. The initial speed is too high for the approximation of a
weak shock, suggesting a medium-intensity shock.

4. \textit{Shock of a moderate intensity}. This intermediate case
is difficult to calculate. The curves in Figure~\ref{fig_wave1}
imply that the wave is initially a rather strong one, and, after
some transitional stage, it becomes weak.

5. \textit{Large-amplitude simple wave} (without discontinuity).
We calculated this case for a uniform density distribution
formally using a solution for a simple wave and a spherical wave
front. The simple wave has a distance-time plot close to the weak
shock fit in Figure~\ref{fig_wave1}; its velocity is less at short
distances. However, the solution of a simple wave rapidly becomes
incorrect. Estimations show that the shock discontinuity appears
either during formation stage of the wave disturbance, or just
afterwards. To fit the calculated distance-time plot to
experimental data points in Figure~\ref{fig_wave1}, the maximum
Lagrangian velocity must be a few times higher than
$v_\mathrm{f}$. The appearance of a type II burst as early as at
00:16\,--\,00:17 was probably due to the rapid steepening the wave
into a shock.

All these curves match more or less observations. Data available
do not to permit us to find out a particular type of a wave. The
spherical strong shock in decreasing density matches observations
better and seems to apply at moderate distances. We therefore use
it in our subsequent discussion of the Moreton wave.

The fact that the best fit for the Moreton and `EIT' waves
corresponds to $\alpha = 2$ is probably accidental. Our idealized
assumptions certainly affect results. We assumed a strong blast
wave and a radial density falloff from the explosion center, which
in reality might be anisotropic and probably variable. The shape
of the wave front is also important; the pile-up mass grows with
distance in a spherical wave faster than in a cylindrical or a
flat wave.

Some range of parameters is expectable in observations. Warmuth
\textit{et al.} (2001, 2004a, 2004b) found average power-law
indices $\delta$ for Moreton waves' distance-time plots of
0.57\,--\,0.62 corresponding to density falloffs of $\alpha =
1.49-1.77$ for spherical waves. The whole range of the measured
indices was mainly from 0.41 to 0.91 ($\alpha = 0.12-2.8$) with
two exceptions of 0.29 and 0.34. Both last events were located on
the limb that made correct identification of wave fronts
difficult; however, such strong damping and deceleration were, in
principle, possible if explosions occurred in media with uniform
horizontal density distributions.

The power-law distance-time relations found by Warmuth \textit{et
al.} (2001, 2004a, 2004b) and in our event are expectable for
coronal blast waves, in accord with a result of Balasubramaniam,
Pevtsov, and Neidig (2007) and the original idea of Uchida (1968)
in a general form. Excitation of a filament by a Moreton wave from
its upper edge in the 6 December 2006 event also could be due to
an incident coronal front inclined down (Gilbert \textit{et al.},
2007). The nature of Moreton waves due to self-similar coronal
shocks is supported by linear growths of wave front thicknesses
with distances found by Warmuth \textit{et al.} (2001, 2004b,
2005) in some events. It is also supported by correspondence with
LASCO data.

\section{Discussion and Conclusion}
\label{Discussion}

\subsection{Overall Scenario of the Event}

The event started with a steady rise of a filament in the active
region. Then an explosive eruption occurred probably due to
mergence of two close filaments. Uralov \textit{et al.} (2002)
argued explosive development of MHD instability in such situation;
Hansen, Tripathi, and Bellan (2004) confirmed this experimentally.
The eruption was likely driven by MHD forces, because the
filaments remained cool. The fastest observed feature with a POS
acceleration $>8$~km\thinspace s$^{-2}$ was bright, which means
that, at least, its parts were at coronal temperatures.

The initial energy of the ejecta was partially spent to disperse
the filaments over a large area. The remainder energy was
insufficient for all fragments to overcome gravity; both escaping
and returning features were observed. The importance of gravity is
manifest in deceleration of all observed features. Since the
ejecta disintegrated, the CME did not resemble a classical
three-part structure with a self-similar frontal structure,
cavity, and core. One part of the ejecta escaped as the CME, and
another one fell back onto the Sun. Masses of both parts were
close. The latter part consisted of fragments of filament(s)
dispersed into a cloud covering almost the whole NW quadrant. The
cloud absorbed background emissions that was observed as moving
EUV dimmings and a `negative radio burst'. The cloud later fell
back onto the solar surface, slipping along magnetic field lines.
Other fragments produced surges visible in H$\alpha$ up to heights
of $\sim\,$200 Mm. The initial speeds of the fastest parts of
filaments were less than the speed of their bright envelope ($\sim
500$~km\thinspace s$^{-1}$ in the plane of sky) by 30--50\%;
fragments of their legs moved still slower. The cooler material
after the explosion was affected by gravity only, had not exceeded
the escape velocity (618~km\thinspace s$^{-1}$), and mainly fell
back. All observed returning features were cool, while no falling
hotter material was seen. By contrast, all observed escaping
features were significantly hotter, at about coronal temperatures,
and just they had counterparts in LASCO images. Since masses of
escaping and falling parts were comparable, the CME was mainly at
$\sim$~1~MK, while the returned part was mostly cool, at $\sim
10^4$~K.

The explosion produced a decelerating coronal blast shock. The
presence of a wave is demonstrated by a displacement of loop
\textit{L1} in Figure~\ref{TRACE}i,j, when the ejecta was far from
it. The skirt of the shock was seen as a Moreton wave decelerating
from the initial speed of $> 1000$ to $\simeq 600$~km\thinspace
s$^{-1}$, while it was observed. At longer distances, the shock
was observed as an `EIT wave' with a speed of $<500$~km\thinspace
s$^{-1}$.

The initial magnetic structure was closed and confined inside the
filament(s). After the eruption, dispersed fragments flew along
lengthy field lines toward the North Pole and landed on a huge
area. Some magnetic structure was carried away by the CME, as
rotation of its component demonstrates. It is difficult to
reconcile these facts without magnetic reconnection, but its role
in the event is not quite clear. Probably, just the destruction of
the magnetic configuration could cause dispersing the ejecta
rather than its ejection as a whole. Anyway, reconnection does not
seem to be directly related with the observed `EIT wave' and
dimmings.

\subsection{Comments on Nature of Large-Scale Disturbances}

The bright oval `EIT wave' in this event was likely due to the
coronal blast shock as follows from its kinematics, which agrees
with the motion of the Moreton wave and theoretically expected
propagation of the shock. The spheroidal shape of the wave front
and correspondence to parameters of the type II radio burst
confirm this conclusion. Quantitative parameters inferred in our
analysis and results of other authors obtained for several
`Moreton \& EIT wave events' show them to be mainly consistent
with their origin due to coronal blast shocks. They are probably
moderate-intensity ones or strong shocks subsequently damping to
moderate waves. Shocks compress (hence brighten) and heat plasmas.
Manifestations of heating were presumably found in observations of
some `EIT waves' (Grechnev \textit{et al.}, 2005; see also
Warmuth, Mann, and Aurass, 2005).

Other large-scale transients of low brightness (visible in
Figure~\ref{running_dif}b north, north-northeast, east-south\-east
of the eruption center, and a sector above the NW limb) moved
slower, had different shapes, and were probably due to a different
reason. On-disk brightenings could be similar to the off-limb
feature. However, they could be due to some processes in low
coronal layers, \textit{i.e.}, different from the off-limb
feature. Their nature is not clear.

Two kinds of dimming were observed. \textit{Shallow dimmings
moving} behind the oval `EIT wave' front were due to absorption of
the Sun's background emission in the moving `cloud' of filament
fragments, as quantitatively confirmed. By contrast, \textit{deep,
long-lived dimmings} \textit{D1}, \textit{D2} in
Figures~\ref{EIT195_fixed_base_dif},\ref{TRACE_dimming} were due
to significant changes of previously bright coronal loops.
Figure~\ref{EIT195_fixed_base_dif} shows that dimming \textit{D2}
was too large to be due to displacement of loops. A plasma density
decrease is the most probable (if not the only) reason that can
occur in two cases. (a)~Previously closed loops open or stretch
into interplanetary space that results in plasma outflows.
(b)~Loops slightly stretch, but do not significantly change. As
can be seen, with the same number of particles in a loop, a change
of its volume from $V_0$ to $V_1$ correspondingly changes emission
measure, $EM_1/EM_0 = V_0/V_1$. Thus, stretch of a loop decreases
its brightness. In summary, deep, long-lived dimmings are likely
due to \textit{density decrease} in coronal structures and
obviously cannot travel. These conclusions agree with those of
Chertok and Grechnev (2005a).

\subsection{Estimations of Mass from Absorption}

We estimated the mass of the ejected `cloud' from absorption in
the radio range, in the $H\alpha$ line, and in EUV lines of
195~\AA\ and 304~\AA. All these methods supplied close results,
from which we conclude that the mass of the `cloud' was $\sim 3
\cdot 10^{15}$~g, a typical mass of a filament. The mass of the
CME was $\sim 2 \cdot 10^{15}$~g, also typical of CMEs. This
implies that such a spectacular event was unlikely an exceptional
one. On the other hand, consistency of the estimates obtained by
means of these methods demonstrates that they offer a promising
way to evaluate masses of on-disk ejecta.

Each of these methods has advantages and limitations. H$\alpha$
images distinctly show absorbing features, but estimations are
constrained to low temperatures and velocities and need unknown
geometrical depths. Estimations from 195~\AA\ emission are not
velocity dependent, but they do not say much about the temperature
of the absorber and fail at low heights. Estimations from the
304~\AA\ emission are sensitive up to the transition region, but
crucially depend on temperature and absorption mechanism (velocity
dependence is important for the resonant scattering). Estimations
from multi-frequency radio data need unknown geometrical depth and
fail at low heights, but they allow to estimate the temperature of
ejecta in a wide range and do not depend on velocity.

\subsection{Summary and Concluding Remarks}
\label{Summary}

From multi-spectral data, an overall picture of the 13 July 2004
eruptive event was reconstructed and confirmed by quantitative
estimations. An explosive filament eruption occurred in an active
region. The ejecta disintegrated into two parts of comparable
masses, one of which flew away as a decelerating CME, and another
part dispersed over almost the whole NW quadrant of the visible
solar disk. The latter part absorbed background solar emission
that was observed as widespread faint moving dimmings at 195~\AA,
a `negative' radio burst, and a huge dimming at 304~\AA. By
contrast, deep, quasi-stationary dimmings also observed at
195~\AA\ in this event were likely due to density decrease in
coronal structures.

Properties of both the Moreton wave and the oval faint front
observed at 195~\AA\ in this event are consistent with their
origin due to a coronal blast shock produced by the eruption. Our
results show what could be reasons for `EIT waves' and dimmings in
a particular event, not pretending to apply to all events.

The coronal blast shock, the Moreton wave, and a huge cloud of the
dispersed material of filaments were interconnected in our event.
It was remarkable, but probably not an exceptional one. CMEs
produced in such events are not expected to accelerate. A
polynomial fit is inappropriate to analyze their kinematics.

Important information about eruptive phenomena could be obtained
using the following methods. Radio observations in a range of
1--10~GHz can supply parameters of ejecta (multi-frequency imaging
observations would be especially valuable). Space-borne 304~\AA\
and H$\alpha$ observations with an interval of $\sim 1$~min would
be important. Masses of on-disk ejecta can be estimated from
absorption.

\acknowledgements

We thank V.~Yurchyshyn for supplying us with BBSO H$\alpha$ data
and I.L. Beigman, M.A. Livshits, S.A. Bogachev, S.M. White, F.
Auch\`ere, G.V. Rudenko, H.R. Gilbert, S. Pohjolainen, and B.
Kliem for useful discussions. We thank the anonymous reviewer for
valuable remarks.

We used the CME catalog generated and maintained at the CDAW Data
Center by NASA and The Catholic University of America in
cooperation with the Naval Research Laboratory. SOHO is a project
of international cooperation between ESA and NASA. We also used
data of the Big Bear Solar Observatory/New Jersey Institute of
Technology, the Mauna Loa Solar Observatory/High Altitude
Observatory, the Nobeyama Solar Facilities, the USAF RSTN Radio
Solar Telescope Network, and the GOES satellites.

The present study is supported by the Russian Foundation of Basic
Research (grants 05-02-17415, 06-02-16106, 06-02-16239,
06-02-16295, and 07-02-00101), the Federal Ministry of Education
and Science (grant 8499.2006.2), and the programs of the Russian
Academy of Sciences `Solar Activity and Physical Processes in the
Sun-Earth System' and `Plasma Heliophysics'.

\end{article}

\end{document}